\documentclass[pra,twocolumn]{revtex4-1}
\usepackage{graphicx}
\usepackage{amssymb}
\usepackage{bm}
\usepackage{amsmath,amsfonts,latexsym,braket}

\usepackage{hyperref}              
\usepackage{comment}
\usepackage{float}
\usepackage{lipsum}                     
\usepackage{xargs}                      
\usepackage[pdftex,dvipsnames]{xcolor}  
\usepackage[colorinlistoftodos,prependcaption,textsize=large]{todonotes}
\newcommandx{\improvement}[2][1=]{\todo[linecolor=Plum,backgroundcolor=Plum!25,bordercolor=Plum,#1]{#2}}

\newcommand*{\br}{\mathbf{r}}

\newcommand*{\bv}{\mathbf{v}}
\newcommand*{\bA}{\mathbf{A}}
\newcommand*{\bR}{\mathbf{R}}
\newcommand*{\bS}{\mathbf{S}}
\newcommand*{\bl}{\mathbf{l}}

\newcommand*{\cO}{{\mathcal O}}

\newcommand*{\cR}{{\mathcal R}}

\newcommand{\subfigimg}[3][,]{%
  \setbox1=\hbox{\includegraphics[#1]{#3}}
  \leavevmode\rlap{\usebox1}
  \rlap{\hspace*{0pt}\raisebox{\dimexpr\ht1-2\baselineskip}{#2}}
  \phantom{\usebox1}
  }

\begin{document}

\title{Infinite Lattices of Vortex Molecules in Rabi-Coupled Condensates}
\author{B. Mencia Uranga}
\author{Austen Lamacraft}
\affiliation{TCM Group, Cavendish Laboratory, University of Cambridge, J. J. Thomson Ave., Cambridge CB3 0HE, UK}
\date{\today}
\email{bm485@cam.ac.uk}

\begin{abstract}
Vortex molecules can form in a two component superfluid when a Rabi field drives transitions between the two components.
We study the ground state of an infinite system of vortex molecules in 2D, using a numerical scheme which makes no use of
the lowest Landau level approximation. We find the ground state lattice geometry for different values of inter-component
interactions and strength of the Rabi field. In the limit of large field when molecules are tightly bound, we develop a
complimentary analytical description. The energy governing the alignment of molecules on a triangular lattice is found to
correspond to that of an infinite system of 2D quadrupoles, which may be written in terms of an elliptic
function  $\mathcal{Q}(z_{ij};\omega_1 , \omega_2 )$. This allows for a numerical evaluation of the energy which enables us to find the ground state configuration of the molecules.

\end{abstract}

\maketitle

\section{Introduction}

Quantized vortices have long been understood to be a characteristic of superfluid flow. Building on Onsager's 1948 announcement of circulation quantization in superfluids \cite{Onsager:1949aa},
and London's work on flux quantization in superconductors \cite{London:1950aa}, the idea that vortex lines form a 2D lattice seems to date from Feynman's 1955 work \cite{Feynman:1955aa}.
Two years later, Abrikosov gave a quantitative theory of the vortex lattice in Type-II superconductors \cite{Abrikosov:1957aa}\footnote{In his Nobel Prize lecture,
Abrikosov states that his discovery dates from 1953 but publication was delayed by an incredulous Landau!}. Experimental confirmation of these predictions arrived shortly afterwards \cite{Vinen:1958aa,Deaver-Jr:1961aa}.

In superfluids and superconductors, vortices form simple, usually triangular lattices. Tkachenko \cite{Tkachenko:1966} proved that for an infinite system of point vortices the triangular lattice has the lowest energy,
and a numerical search of up to 11 vortices per unit cell within the same model found no other stable configurations \cite{Campbell:1989aa}.
Kleiner \emph{et al} \cite{Kleiner} showed that in the opposite limit of very large vortices (relative to separation), the infinite lattice orders in a
triangular geometry (they show that this state has a lower energy than the square lattice that Abrikosov
had erroneously suggested as a ground state \cite{Abrikosov:1957aa}). Brandt later showed that the stability of the triangular lattice persists through the entire range of vortex sizes \cite{Brandt:1972}.
Are more complicated crystal structures possible? Superfluids with multicomponent order parameters,
where vortices may form in different components, provide one avenue. Historically, the first such superfluid was $^{3}$He \cite{Salomaa:1987}, while atomic Bose condensates with internal spin states are a second, more recent example \cite{Stamper-Kurn:2013aa}.

As in the solid state, one route to more complicated structures is to decorate a crystal structure with `molecules` made of two or more vortices. This is the situation that will concern us. In 2002 Son and Stephanov \cite{Son-Stephanov} predicted the existence of a \emph{vortex molecule}  in a Rabi-coupled two component condensate. Subsequent works have focused on the dynamics of a single molecule \cite{Tylutki:2016aa,Qu:2017aa,Calderaro}, as well as ground state  properties  \cite{Garcia-Ripoll:2002,Kasamatsu:2004aa}. Lattices of vortex
molecules in a harmonic trap were studied in Refs.\onlinecite{Cipriani:2013aa,Aftalion:2016aa}.

To the best of our knowledge, vortex molecules have not been experimentally
observed. There is a different object, sometimes also called vortex molecule \cite{Salomaa:1985aa}, which has recently been observed in the
polar phase of superfluid $^{3}$He \cite{Autti:2016aa}.
The observed object is made of two half-quantum vortices ($\pi$ winding of the phase), whereas the vortex molecule we are concerned with is made of two integer vortices ($2\pi$ winding of the phase). They share the feature that the vortices are linked by a domain wall which leads to confinement of the pair, although the repulsion that balances the tension in the domain wall has a different origin, as we explain in Section \ref{sec:mol}.

Infinite vortex lattices have previously been studied both within the Lowest Landau Level (LLL) approximation and beyond
\cite{Ho:2001,Cooper:2008aa}, both for single component \cite{Abrikosov:1957aa,Kleiner,Luca} as well as multicomponent
condensates \cite{Kita:2002,Reijnders:2004,Ho:2002,Keceli:2006}.
For further work on vortex lattices, see the reviews \cite{Cooper:2008aa,Fetter:2009}.

This paper concerns the structure of \emph{infinite} arrays of vortex molecules in 2D. To orient our discussion, the remainder of the introduction introduces the theoretical model and describes the physics of a single vortex molecule, before we move on to the case of a lattice.

\subsection{Hamiltonian}

We consider an infinitely extended, rotating two component spinor Bose-Einstein Condensate in 2D. In equilibrium, the thermodynamic quantity to minimize is the free energy (or energy at $T=0$) in the rotating frame. In the presence of an AC field that gives rise to Rabi oscillations, the relevant low energy Hamiltonian is $H=H_0 + H_{\text{int}} + H_{\text{Rabi}}$, where
\begin{subequations}
\begin{align}
H_0=&  \sum_{\sigma}\int d\mathbf{r} \ \Psi_{\sigma}^{\dagger}(\mathbf{r}) \left[ \frac{\mathbf{p}^2}{2} + \frac{\omega^2 r^2}{2} - \mathbf{\Omega} \cdot \mathbf{L} \right]  \Psi_{\sigma}(\mathbf{r}) \nonumber \\
=&  \sum_{\sigma}\int d\mathbf{r} \ \Psi_{\sigma}^{\dagger}(\mathbf{r})  \left[ \frac{(\mathbf{p} - \mathbf{A})^2}{2} + \frac{\omega_{\text{eff}}^{2} r^{2}}{2}  \right]  \Psi_{\sigma}(\mathbf{r}) , \\
H_{\text{int}}=&  \sum_{\sigma_{1}\sigma_{2}} \frac{g_{\sigma_1 \sigma_2}}{2}
\int d\mathbf{r}  \  \Psi_{\sigma_{1}}^{\dagger}(\mathbf{r})\Psi_{\sigma_{2}}^{\dagger}(\mathbf{r}) \Psi_{\sigma_{2}}(\mathbf{r})\Psi_{\sigma_{1}}(\mathbf{r}), \\
\label{H_Rabi}
H_{\text{Rabi}}=& - \Omega_R \int d\mathbf{r} \left[ \Psi_{a}^{\dagger}(\mathbf{r})\Psi_{b}(\mathbf{r}) + \Psi_{b}^{\dagger}(\mathbf{r})\Psi_{a}(\mathbf{r}) \right].
\end{align}
\end{subequations}
($\hbar = m = 1$) Here, the operators $\Psi_{\sigma}^{\dagger}(\mathbf{r})$ create bosons at position $\br$
with spin $\sigma$, $\mathbf{A}\equiv\mathbf{\Omega} \times \br$, $\omega_{\text{eff}}\equiv\sqrt{\omega^2 - \Omega^2}$, $\omega$ is the harmonic trap frequency,
$\mathbf{\Omega}\equiv \Omega \hat{z}$ is the angular velocity of the trap, $\mathbf{L}$ is the angular momentum operator. The dimensionless couplings $g_{\sigma_1 \sigma_2}>0$ are
the strength of the hyperfine state dependent interatomic contact interactions and $\Omega_\text{R}$ is the
Rabi frequency. The external electromagnetic field is introduced in the dipole approximation through $H_{\text{Rabi}}$ \cite{quantum-optics}.

\subsection{Single Vortex Molecules}\label{sec:mol}

In their seminal work, Son and Stephanov predicted that in a Rabi-coupled 3D two component BEC,  there should exist a domain wall of the relative phase of the two components --a domain wall inside which the
relative phase changes by $2 \pi$ \cite{Son-Stephanov}. This would be bound by a closed vortex line. Furthermore, they argued that the external field would work as a confinement mechanism for vortices of different components.

In this section we give qualitative arguments to explain why in 2D a pair of such vortices are confined in a \emph{vortex molecule}. In the mean field treatment discussed in Section \ref{sec:gp}, $H_\text{int}$ and $H_\text{Rabi}$ give rise to contributions
\begin{align}
&E_{\text{int}} = \frac{g}{2}   \int d \mathbf{r} \  \left[\rho_a(\br)+\rho_b(\br)\right]^2 \; \text{and} \\
\label{E_Rabi}
&E_{\text{Rabi}} = -2 \Omega_\text{R} \int d \mathbf{r} \  \sqrt{\rho_{a}(\mathbf{r}) \rho_{b}(\mathbf{r})} \cos(\theta_{a}(\mathbf{r})-\theta_{b}(\mathbf{r})).
\end{align}
Here we have set $g\equiv g_{aa}=g_{bb}=g_{ab}$ for simplicity, and have introduced the amplitude-phase (Madelung) representation $\psi_a(\br) =\sqrt{\rho_a(\br)}e^{i\theta(\br)}$.

For $g>0$, $E_{\text{int}}$ favors configurations where densities of different components don't overlap. $E_{\text{Rabi}}$ favors alignment of
the phases of the two components. Now let's think of a pair of vortices, one in each component. In the absence of the external field, there is
no energy cost for having the phases misaligned:

\begin{figure}[H]
\centering
\includegraphics[width=8cm]{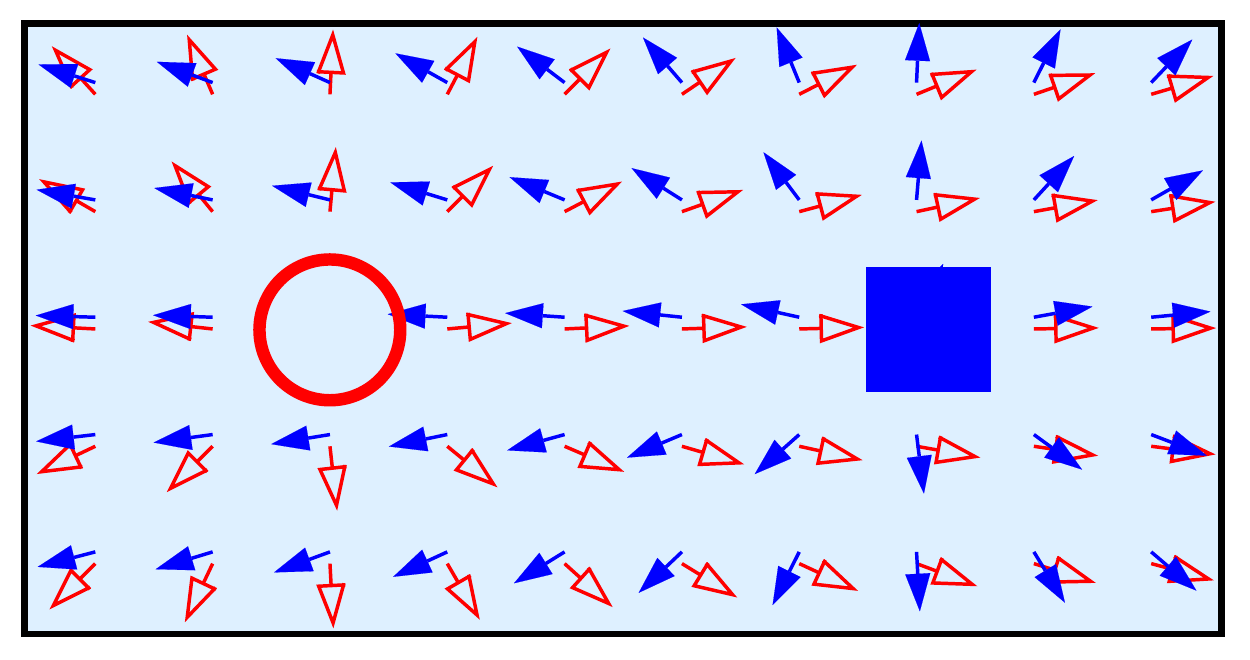}
\caption{Schematic view of a pair of free vortices, one in each component, in the absence of an external field. There is no energy cost for having phases misaligned and the winding is smooth to minimize the
kinetic energy.}
\label{fig:1}
\end{figure}
The term $E_{\text{Rabi}}$ is minimized by full alignment of the phases.
This is achieved if the two vortices overlap completely. On the other hand, overlapping vortices have an increased
 $E_{\text{int}}$ relative to nonoverlapping vortices. Thus there is a competition between $E_{\text{Rabi}}$ and $E_{\text{int}}$, which have typical magnitudes per particle of $\Omega_\text{R}$ and $g n$, where $n$ is the bulk density of the two components (assumed equal).

 Between the limits $\Omega_\text{R} \ll g n$ and $\Omega_\text{R} \gg g n$ the optimal arrangement will be a configuration where vortices are neither overlapping, nor too far separated.

\begin{figure}[H]
\centering
\includegraphics[width=8cm]{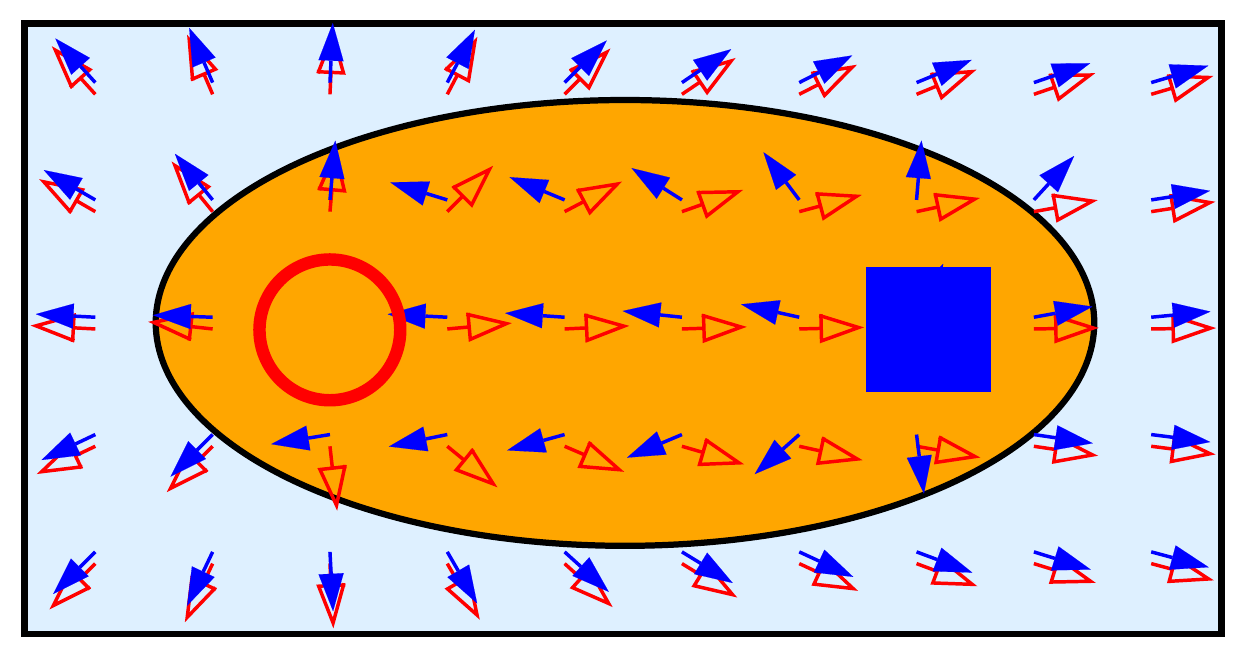}
\caption{Vortex pair in the presence of the external field. Because of the energy cost for misaligning phases
Eq.~\eqref{E_Rabi}, the misalignment region (in orange) is confined in a region. Out of this region the phases are aligned.}
\label{fig:2}
\end{figure}
Hence in the presence of vortices, which is ensured by the rotation of the trap (above some critical angular velocity $\Omega_c$), the external field $\Omega_\text{R}$ works as an inter-component \emph{confinment  mechanism} for vortices,
giving rise to vortex molecules. The size of the molecule (expressed in terms of the healing length $\xi = (4gn)^{-1/2}$) is a function of the ratio $gn/\Omega_\text{R}$, but there appears to be no simple argument for this relationship. In $^{3}$He-A the situation is different: the equilibrium separation of a molecule arises from a balance between tension in the domain wall and the logarithmic repulsion of the vortices \cite{Salomaa:1985aa}. In our setting the vortices are in different components, and the repulsion arises from $E_{\text{int}}$. Thus there is not a simple `phase only' description of the molecule.

\subsection{Outline}

The outline of the remainder of the paper is as follows. In Section II we introduce the method we use to find the ground state and show the results we obtain. In Section III we introduce an effective theory in terms
electric point charges for the system of vortex molecules and we show the ground state configuration. Section IV contains the conclusions. Some of the details of the calculations are described in Appendices A-D.

\section{Gross--Pitaevskii Theory}\label{sec:gp}

We want to study the ground state properties of the above mentioned Spinor BEC. Gross-Pitaevskii theory describes the properties of the condensate at $T=0$.
The approach used in this theory is variational, i.e. we have an interacting Hamiltonian whose exact ground state we don't know and we use our experimental knowledge about the existence of a condensate at $T=0$, to guess the ground state wave function.

In a two level system with off-diagonal coupling, the eigenstates are superpositions of the eigenstates of the Hamiltonian in the absence of coupling  \cite{cohen1998quantum,Kasamatsu:2004aa,Matthews}.
The off-diagonal coupling mediated by the external electromagnetic field Eq.~\eqref{H_Rabi}, motivates an ansatz where there is a $N$ particle condensate in a state which
is a superposition of two hyperfine states $\{ | \phi_{a} \rangle,| \phi_{b} \rangle \}$ :

\begin{equation}
| \Psi \rangle =\frac{1}{(N!)^{1/2}} \left (  \sum_{\sigma} \int  d \mathbf{r} \ \phi_{\sigma}^{*}(\mathbf{r})  \Psi^{\dagger}_{\sigma}(\mathbf{r}) \right )^N  | 0 \rangle .
\end{equation}
We then calculate the expectation value of the Hamiltonian in this ansatz state.
Defining the condensate wave-function $\psi_{\sigma}(\mathbf{r}) = \sqrt{N} \phi_{\sigma}(\mathbf{r})$ and using that $N(N-1) \sim N^2$, the final expression to be minimized is:

\begin{align}
\label{E}
E(\psi_{a}, \psi_{b}) &=  \sum_{\sigma}\int d\mathbf{r} \ \psi_{\sigma}^{*}(\mathbf{r})  \left[ \frac{(\mathbf{p} - \mathbf{A})^2}{2} + \frac{\omega_{\text{eff}}^{2} r^{2}}{2} \right]  \psi_{\sigma}(\mathbf{r}) \nonumber \\
&+ \sum_{\sigma_1  \sigma_2} \frac{g_{\sigma_1 \sigma_2}}{2}  \int d \mathbf{r} \  |\psi_{\sigma_1}(\mathbf{r})|^2 |\psi_{\sigma_2}(\mathbf{r})|^2  \nonumber \\
&-\Omega_\text{R} \int d \mathbf{r} \left[\psi_{a}^{*}(\mathbf{r}) \psi_{b}(\mathbf{r}) +\psi_{b}^{*}(\mathbf{r}) \psi_{a}(\mathbf{r})\right].
\end{align}

\subsection{Infinite Lattices}

In order to study the infinite lattice, we choose $\omega_{\text{eff}}=0$. It is the value of the effective trapping potential $\omega_{\text{eff}}$ that governs the ``envelope'' modulation of the condensate density. In the
absence of an effective trapping potential, the only modulation is the one due to the presence of a vector potential $\bA$. Hence, $\omega_{\text{eff}}=0$ corresponds to having a spatially extended condensate, where there are no boundary effects and an ideal vortex lattice is expected to be found in the ground state \cite{Luca}. We stress that periodicity -- which we assume from now on -- is not \emph{a priori} obvious.

It is important to comment on the difference between the \emph{unit  cell} of the lattice and what we call the \emph{computational unit cell}. The computational unit cell is the system in which we do the calculations: for
numerical simplicity we choose a rectangular system. Note that this does not imply that the unit cell (the smallest portion of the system that repeats in the infinite periodic system) should be rectangular.
The easiest example that illustrates this subtlety well is the triangular lattice. The unit cell is a rhombus and contains one vortex. On the other hand, if one is restricted to have a rectangular computational unit cell, this would
be a rectangle with aspect ratio $\cR=1/ \sqrt{3}$ (or $\sqrt{3}$) containing two vortices. Both unit cells reproduce the same infinite lattice.

What conditions should be imposed at the boundary of the computational unit cell? It is most natural to require gauge-invariant quantities to be periodic under some set of translations. In order to fully determine the boundary condition fulfilling the aforementioned condition, it is necessary and sufficient to require
periodicity of densities, velocities and pseudo-spin: $\rho_{\sigma}, \bv_{\sigma}$ and $\bS$ (Appendix~\ref{boundary conditions}). This leads to the boundary conditions \cite{Kita:1998,Kita:2002}:

\begin{align}
\label{BC1}
 &\psi_{\sigma}(x+L_x , y) = e^{i \Omega L_x y} \psi_{\sigma}(x,y), \nonumber \\
 &\psi_{\sigma}(x, y+L_y) = e^{-i \Omega L_y x} \psi_{\sigma}(x,y) , \; \sigma = a,b.
\end{align}
Here $L_x$ and $L_y$ are the dimensions of our rectangular computational unit cell. In order to have a consistent theory, the angular velocity of the trap can only take a
discrete set of values (Appendix~\ref{allowed values and relation to the number of vortices}):

\begin{equation}
\Omega = \pi n_v,
\end{equation}
where $n_v = N_v/L_x L_y$ is the vortex density in the computational unit cell.

Recently Mingarelli \emph{et al} \cite{Luca:2017} have studied infinite vortex lattices in a two component superfluid.
The boundary conditions used in this work allow for non periodic spin solutions (Appendix~\ref{boundary conditions}).

\subsection{Numerical Calculations}

To find the computational unit cell and the associated ground state wavefunction, we numerically minimize the discrete version of Eq.~\eqref{E} (see Eq.~\eqref{E_d}) subject to the constraint of fixed particle number for each component, with $\omega_{\text{eff}}=0$ and using the boundary conditions Eq.~\eqref{BC1}. The method used is the nonlinear conjugate-gradient
algorithm as implemented in SciPy \cite{Jones:2001aa}. As pointed out by Mingarelli \emph{et al}, in order to allow the vortex lattice configuration to access any lattice geometry in the minimization process, the energy Eq.~\eqref{E} has to be minimized not only with respect to the wave-functions but also the aspect ratio $\cR=L_x / L_y$ \cite{Luca,Kita:2002}.

To find the computational unit cell in the ground state, we use the following procedure:

\begin{enumerate}
  \item Minimize the energy for a given $A=L_x L_y$ and $N_v$, to find $E_{min}$, $\cR_{min}$ and
        $\{ \psi_{\sigma,min}\}$.

  \item Repeat the minimization with area $2A$ and $2N_v$ vortices.

  \item If the energy has doubled and $\cR_{min}$ has doubled (and halved --  note that in general there are several $\cR_{min}$ corresponding to one same configuration, at least $\cR_{min}$ and $1/\cR_{min}$), we can infer that the unit cell contains $N_v$ vortices and its aspect ratio is $\cR_{min}$. If either $E_{min}$ was not doubled or $\cR_{min}$ was not doubled and halved, we keep increasing the area and the number of vortices until the \emph{doubling-halving criterion} has been fulfilled.

  \item We repeat the same protocol for several starting $N_v$. We pick the solution with smallest energy density that fulfills the  doubling-halving criterion.
\end{enumerate}
Step 3 follows from the fact that by stacking unit cells together, one should be able to reproduce the infinite lattice. Since $\mathcal{R}=L_x / L_y$, $\mathcal{R}_{min}$ should be doubled if we stack two unit cells horizontally, and halved if we stack them vertically. The judgment of the fulfillment of the doubling-halving criterion, takes into account the integration error Eq.~\eqref{error}. This protocol does not ensure one to find the true ground state. Note that even if we find a choice of $N_v$ that fulfills the doubling-halving criterion, the possibility always exists that there could be a larger $N_v$ with a lower energy density.

Note that when $\Omega_\text{R}=0$ and $g_{aa}=g_{bb}=g_{ab}$, the energy Eq.~\eqref{E} is invariant under unitary transformation:
\begin{equation}
\label{unitary}
 \binom{\psi_{a}}{\psi_{b}} \rightarrow U \binom{\psi_{a}}{\psi_{b}}, \; \text{$U$  unitary. }
\end{equation}
As a consequence there is a continuous manifold -- in fact a sphere -- of ground states related by unitary transformation.

When $\Omega_\text{R}\neq0$ and $g_{aa}=g_{bb}=g_{ab}$, the symmetry of the energy Eq.~\eqref{E} is lowered, but it still is invariant
under rotations of the spinor in the $yz$ plane (note that $E_{\text{Rabi}}$ is the integral of $S_x$ Eq.~\eqref{E_Rabi}):
\begin{equation}
\label{Sx}
 \binom{\psi_{a}}{\psi_{b}} \rightarrow e^{i(\theta/2)\sigma_x} \binom{\psi_{a}}{\psi_{b}}.
\end{equation}
While the densities in each component $\rho_{\sigma}$ are not invariant, the total density is. In terms of $\rho_{a,b}(\br)$, therefore, one can find several vortex lattice geometries in the ground state.

For $\Omega_\text{R}\gg g_{ab}n$ we recover the behaviour of a scalar condensate in the state $\psi_a(\br)=\psi_b(\br)$: the two vortices of each molecule (see Fig.~\ref{fig:2}), overlap. Thus, we recover the triangular lattice geometry, independent of the value of $\alpha \equiv g_{ab}/\sqrt{g_{aa}g_{bb}}$.
We now turn to the finite $\Omega_\text{R}$ behavior. From now on we choose the values
$g_{aa}=g_{bb} \equiv g =0.125$ and $\mu_a = \mu_b \equiv \mu = 12.5 n_v$, and explore three different values of $\alpha$.
From the Euler-Lagrange equations one can deduce that in this case the bulk densities are $n=\frac{\mu + \Omega_R}{g+g_{ab}}$ and the healing lengths
$\xi=\frac{1}{\sqrt{2(\mu + \Omega_R)}}$. The choice of $\xi$ ensures that we are both away from the LLL and point-vortex limits.

\subsubsection{$\alpha=1$}

We begin with the case of zero Rabi field: $\Omega_\text{R} =0$ (Fig.~\ref{fig:3}(a)). For this case we find that the computational unit cell
contains four vortices in each of the components (the unit cell has two) and the vortex lattice is composed of two intertwined
rectangular sublattices. Note that as explained above in Eq.~\eqref{unitary}, this is only one lattice of an infinite degenerate
set.
The two sublattices give rise to an overall (neglecting the two different flavors) triangular lattice $\cR=1/\sqrt{3}$.
Our finding agrees with
\cite{Ho:2002}, even though we are away from the LLL limit. It would be interesting to study this problem in the opposite limit of very small healing length \cite{Lamacraft:2008aa}, to see whether the triangular lattice is the ground state in this case too.

Here it is interesting to note that although the computational unit cell of the density contains two vortices of each component,
the pseudo-spin has a period that is twice as large, that is why the overall
computational unit cell has four vortices.

We then include $\Omega_\text{R}$, which gives rise to molecules. In Fig.~\ref{fig:4}(a) we see that neighbouring
molecules tend to antialign. On the other hand, we find a continuous degeneracy with respect to the alignment direction, i.e. there
is no preferred direction.

\subsubsection{$\alpha=0.5$}

For $\Omega_\text{R}=0$ we find a square chess board configuration of vortices with four vortices per component (Fig.~\ref{fig:3}(b)).
This result agrees qualitatively with \cite{Ho:2002}.

\begin{figure}[t]
\centering
\subfigimg[height=4cm,angle=0]{\raisebox{-115pt}{ \hspace*{20pt} (a)}}{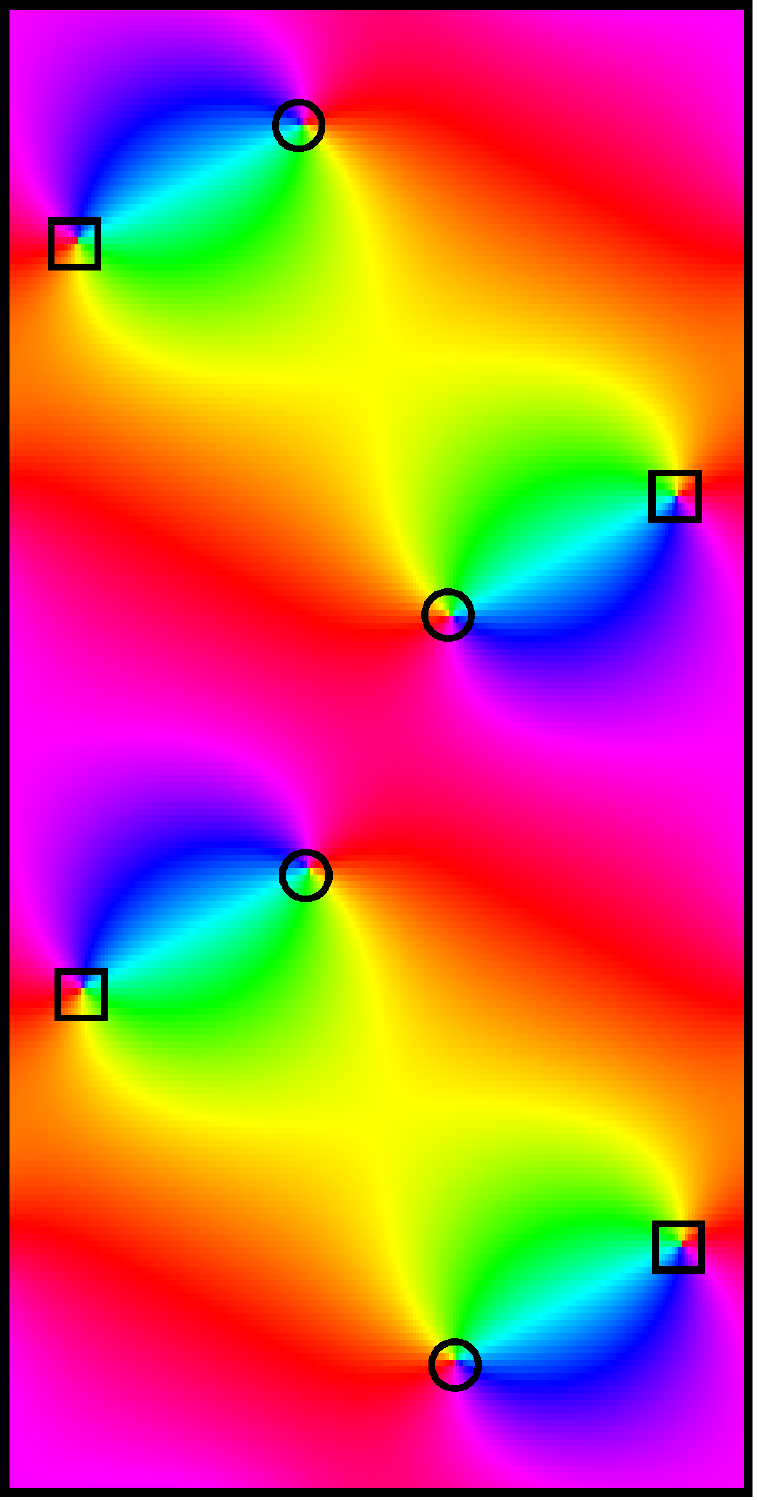}
\subfigimg[height=4cm,angle=0]{\raisebox{-115pt}{\hspace*{30pt} (b)}}{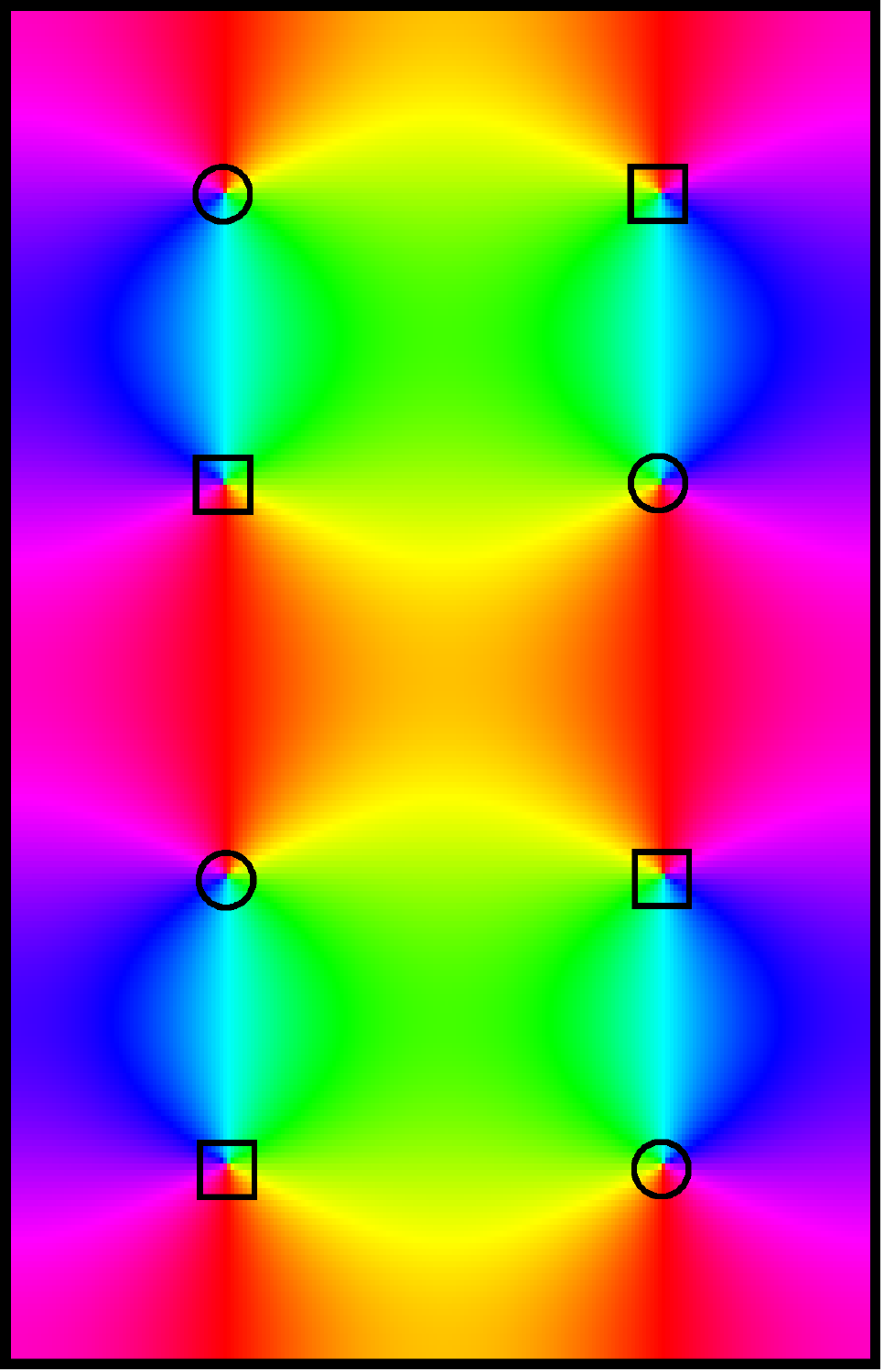}
\subfigimg[height=4cm,angle=0]{\raisebox{-115pt}{\hspace*{35pt} (c)}}{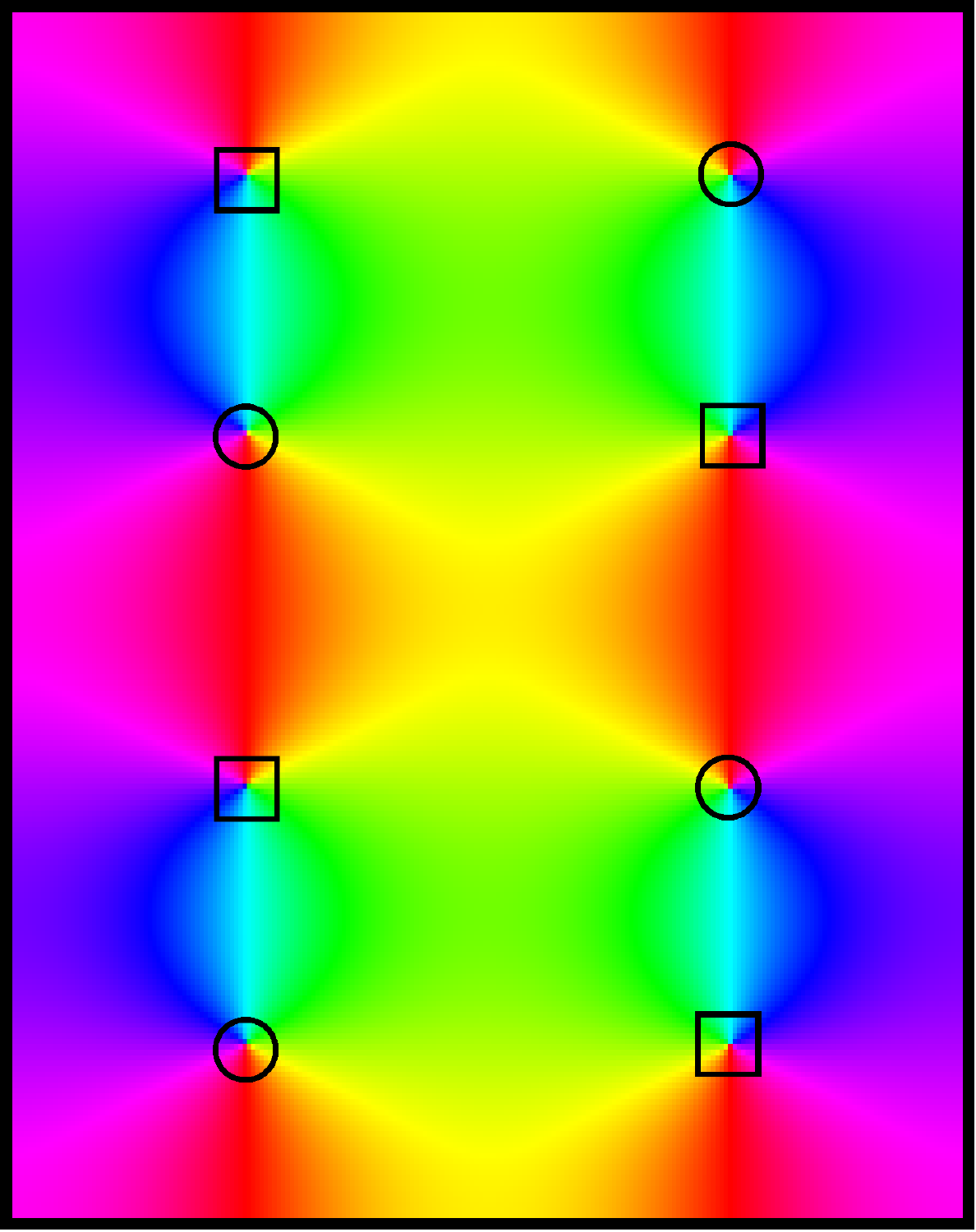}
\hspace*{-2pt}\raisebox{-9.5pt}{\subfigimg[width=0.0373\textwidth]{\raisebox{23pt}{ \hspace*{-3pt} $\varphi$}}{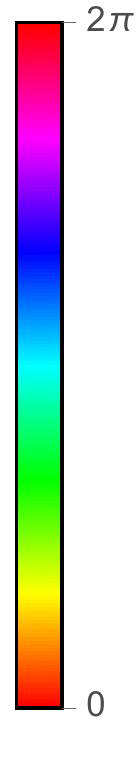}}
\caption{The colorcode shows the relative phase $\varphi=\theta_b-\theta_a$. From left to right, we show the computational
unit cell for $\Omega_\text{R}/n_v = 0.5,0.25,0.1$ and $\alpha = 1, 0.5, 0.2$, respectively. The aspect ratios are
$\cR = 0.5,0.64,0.79$. The ground state aspect ratios are the given ones, only within integration error, i.e.
there is a set of $\cR$ within integration error that fulfill the doubling-halving criterion. Our finding agrees qualitatively with what was
found in \cite{Cipriani:2013aa} (except $\alpha = 0.2$), in the center of the trap.
Note that for $\alpha=1$ repeated computations yield different axes along which molecules are antialigned.}
\label{fig:4}
\end{figure}

For $\Omega_\text{R}\neq 0$, the square chess board gets distorted
giving rise to a rectangular chess board (Fig.~\ref{fig:4}(b)). Again molecules like to align or antialign. In this case, there is a
preferred direction of alignment parallel to one of the Cartesian axes.

\subsubsection{$\alpha=0.2$}

\begin{figure}[b]
\centering
\subfigimg[width=4.5cm,angle=90]{\raisebox{-115pt}{ \hspace*{25pt} (a)}}{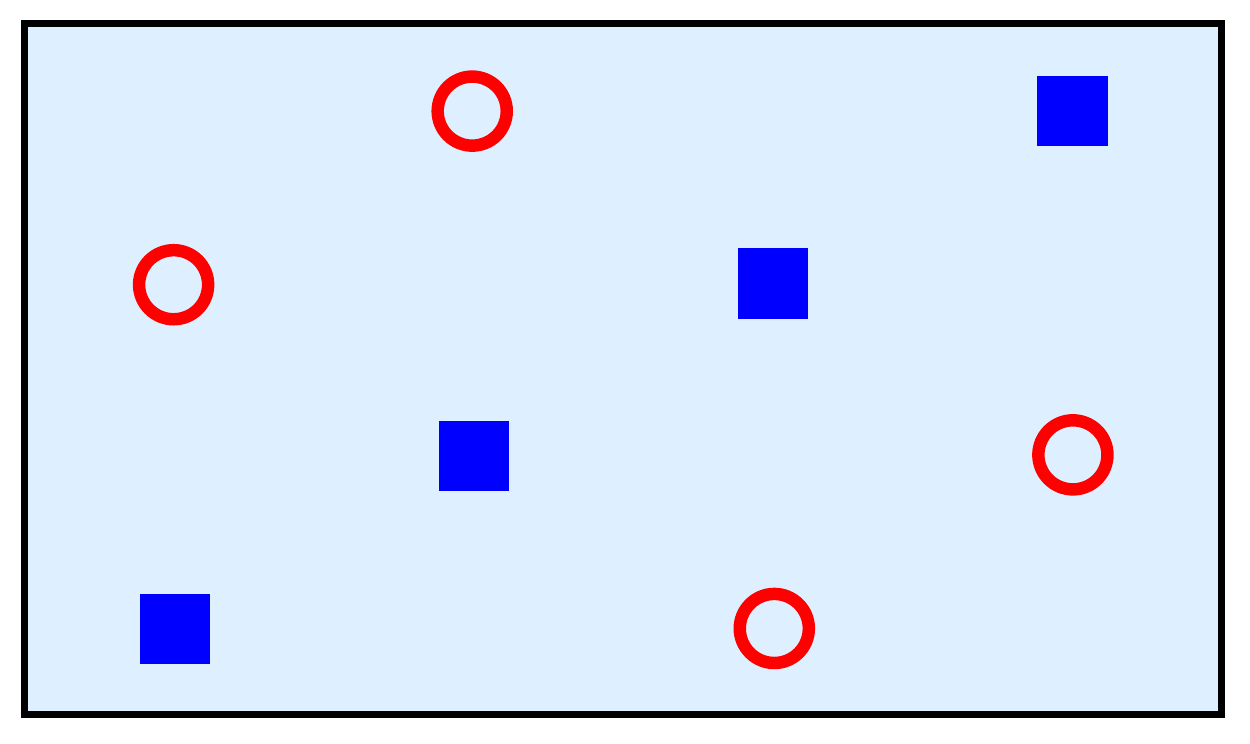}
\subfigimg[width=4.5cm,angle=90]{\raisebox{-115pt}{\hspace*{25pt} (b)}}{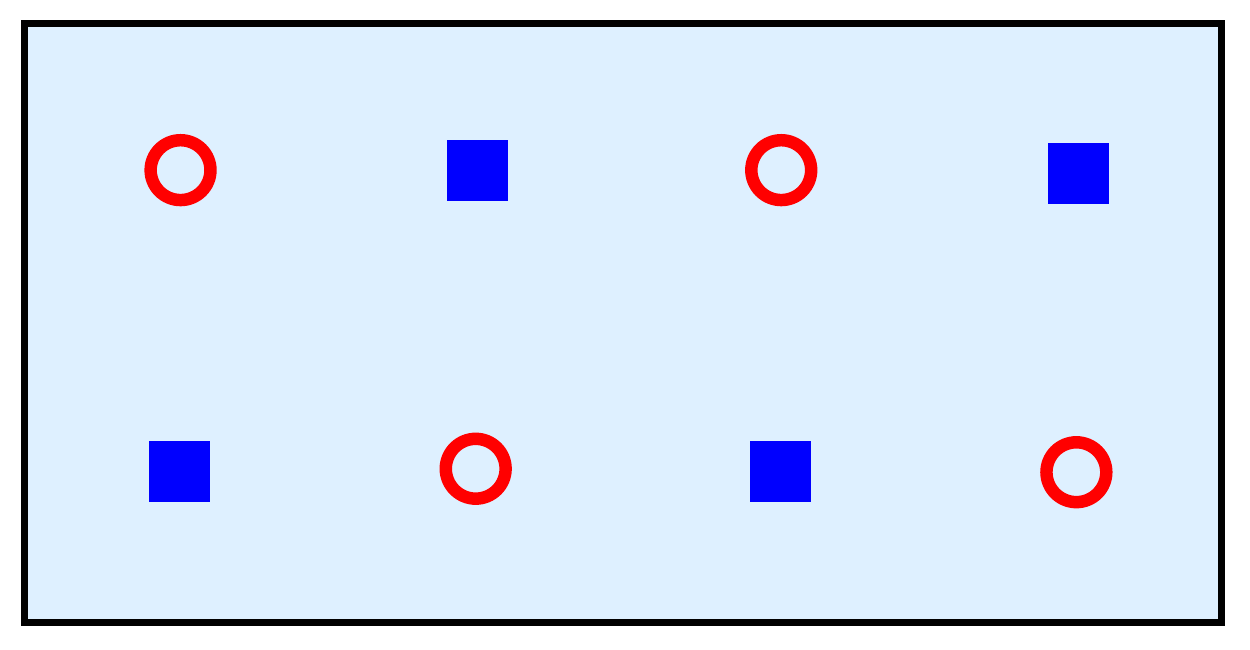}
\subfigimg[width=4.5cm,angle=90]{\raisebox{-115pt}{\hspace*{30pt} (c)}}{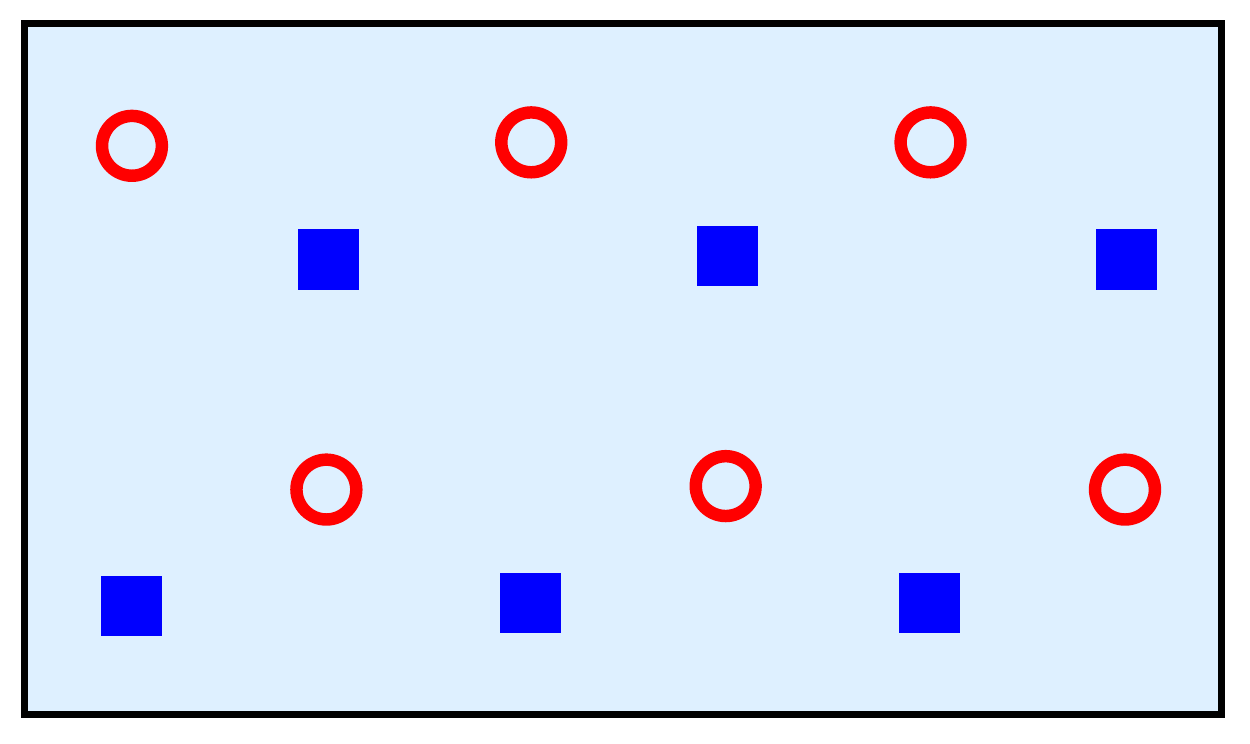}
\caption{Schematic view of the computational unit cell. From left to right, we show the computational unit cell for $\Omega_\text{R}/n_v = 0$ and $\alpha = 1, 0.5, 0.2$,
respectively. The aspect ratios are
$\cR = 1/\sqrt{3},0.5,1/\sqrt{3}$. The ground state aspect ratios are the given ones, only within integration error, i.e.
there is a set of $\cR$ within integration error that fulfill the doubling-halving criterion. Our finding agrees qualitatively with what was
found in \cite{Ho:2002}.}
\label{fig:3}
\end{figure}

For $\Omega_\text{R}=0$ we find two intertwined triangular lattices with six vortices per component in the computational unit cell (Fig.~\ref{fig:3}(c)). This also agrees
with \cite{Ho:2002}.

When we include $\Omega_\text{R}$, the result is qualitatively similar to the result obtained for $\alpha = 0.5$. The geometry of the lattice differs from the one obtained in \cite{Cipriani:2013aa} in the center of the trap. This may be due to either differences in the parameters used, because the computational unit cell is very large (we tried up to 16 vortices per component) or because for the size of the trap (relative to the healing length) used in \cite{Cipriani:2013aa}, is not large enough to obtain the structure of the ideal infinite system in the bulk of the trap.

\subsection{Orientation of the Molecules}

The numerical calculations show that for $\alpha = 1$ there is no preferred direction of alignment of the molecules whereas for
$\alpha < 1$ there is.  To understand this it is helpful to consider the configuration of the pseudo-spin density $\bS = \Psi^\dagger \boldsymbol{\sigma}\Psi / 2$. An isolated molecule has pseudo-spin pointing in the $-x$ direction at its center, and in the $+x$ direction far outside. Moving out from the center, the points where the spin points in the $\pm z$ direction give the location of the vortices in the two components.

Expressing the interaction and Rabi energies in terms of the pseudo-spin density gives
\begin{align}
\label{E_int_Sz}
&E_{\text{int}} = \frac{g}{4}   \int d \mathbf{r} \  \left[4 S_z^2(\br) (1-\alpha) + \rho^2(\br) (1+\alpha)\right] \; \text{and} \\
&E_{\text{Rabi}} = -2\Omega_\text{R} \int d \mathbf{r} \  S_x(\br),
\end{align}
where $\rho = \rho_a + \rho_b$. For $\alpha = 1$ the interaction energy is independent of the spin direction, so configurations that differ by global rotations of the spin around the $x$-axis have the same energy. Such a global rotation causes the two vortices forming the molecule to rotate about their center, explaining the numerical observation that the orientation is undetermined.

\begin{figure}[h]
\centering
\subfigimg[height=4.0cm,angle=0]{\raisebox{-105pt}{ \hspace*{20pt} (a)}}{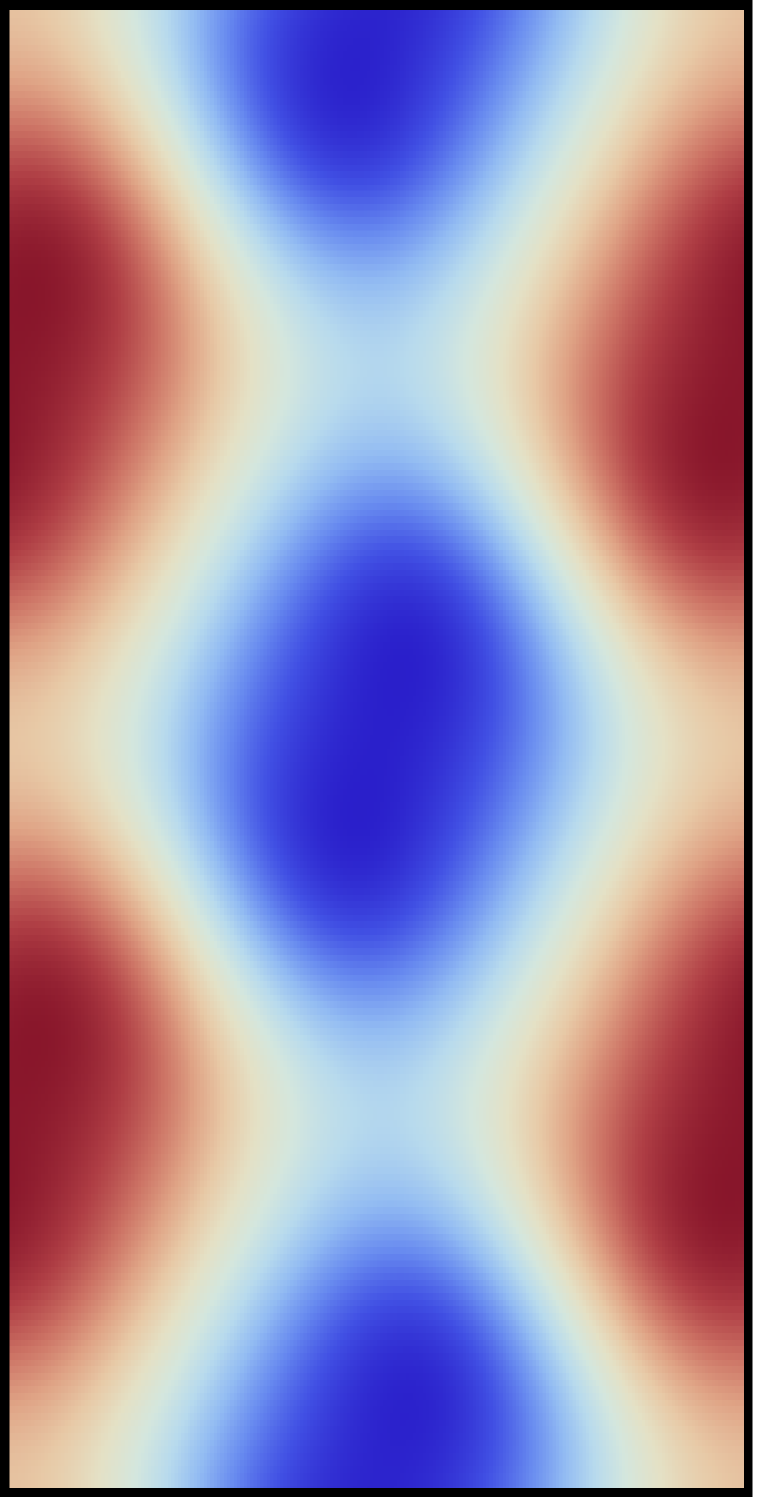}
\subfigimg[height=4.0cm,angle=0]{\raisebox{-105pt}{\hspace*{30pt} (b)}}{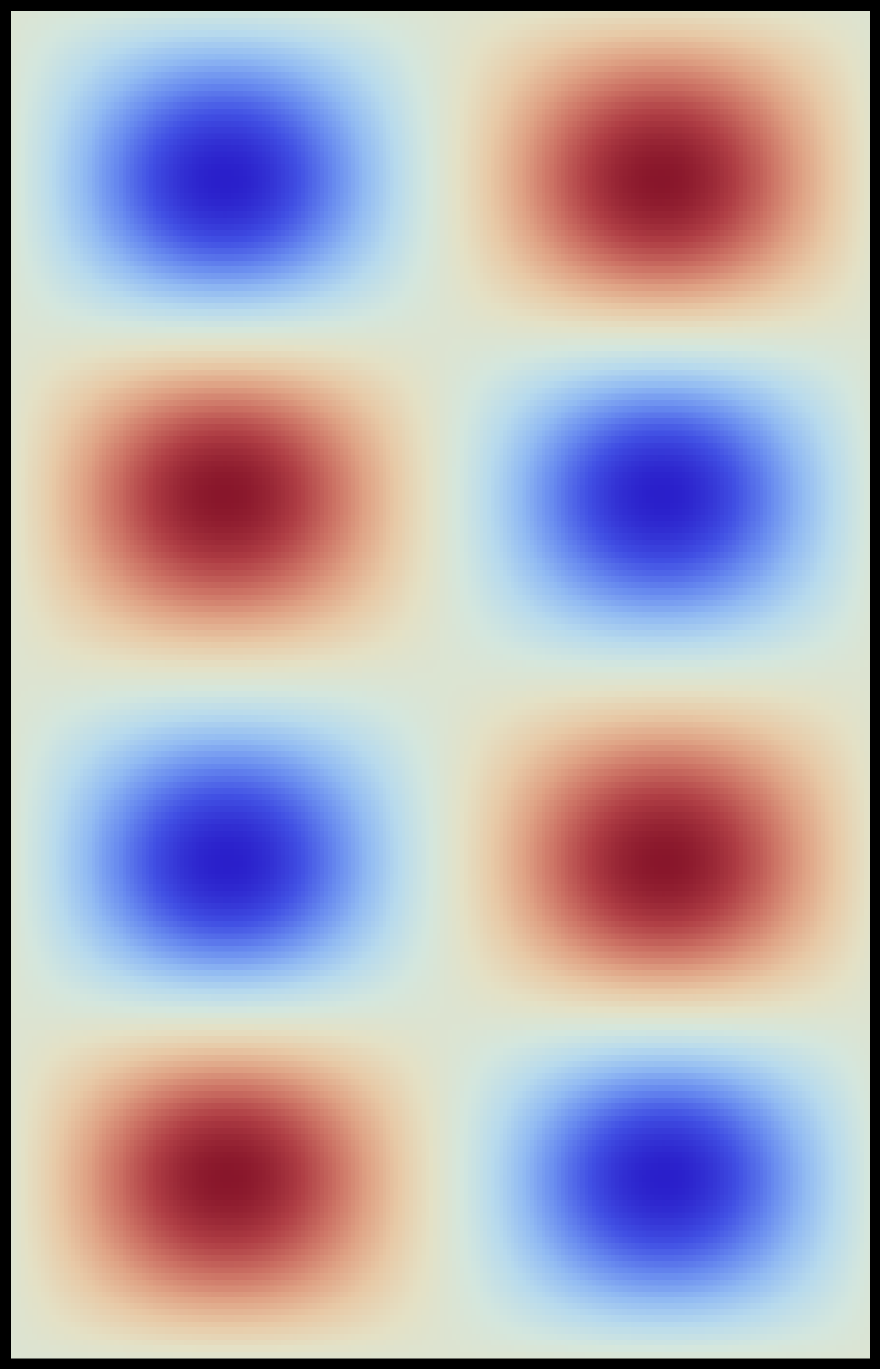}
\subfigimg[height=4.0cm,angle=0]{\raisebox{-105pt}{\hspace*{35pt} (c)}}{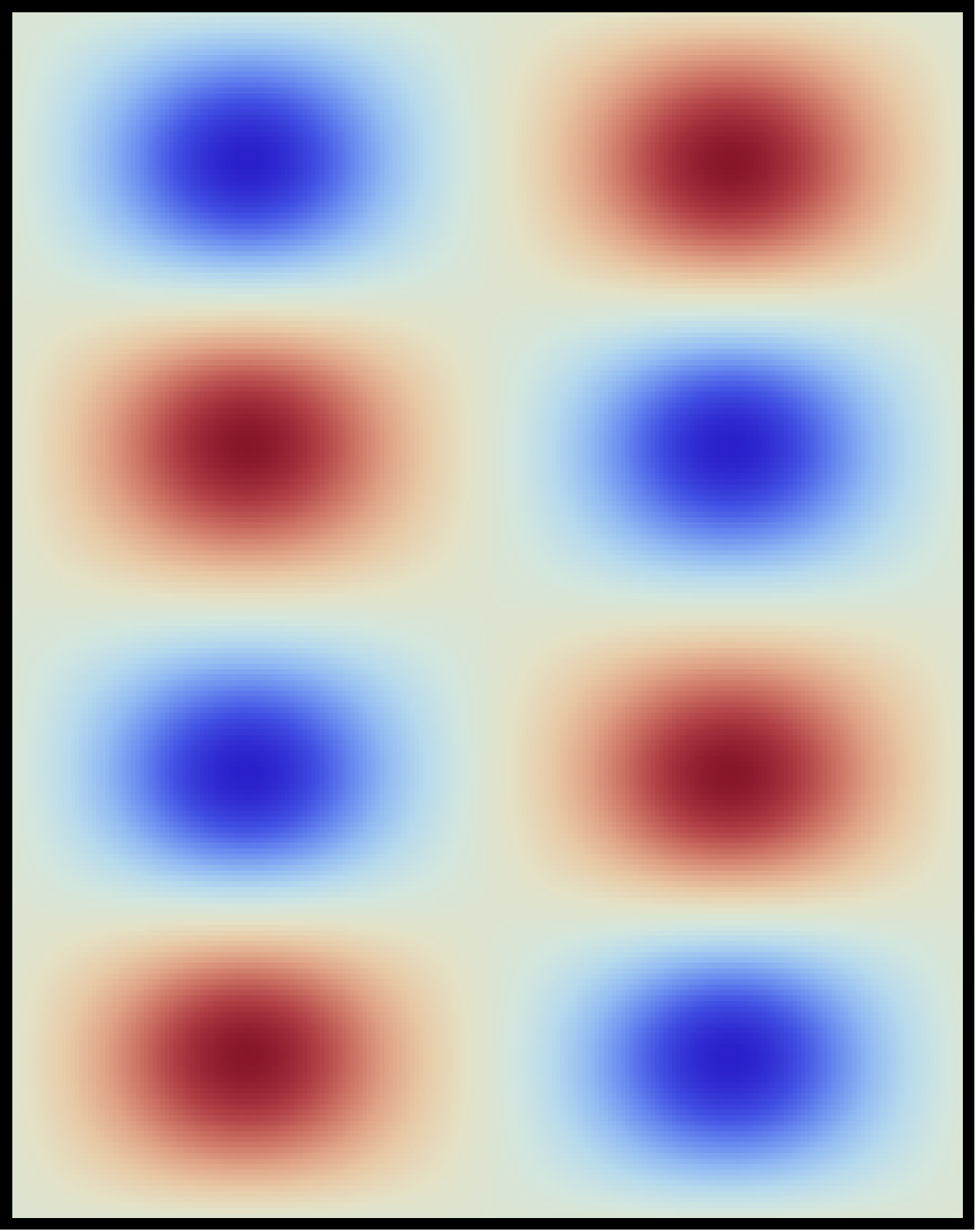}
\hspace*{-2pt}\raisebox{-9.5pt}{\subfigimg[width=0.025\textwidth]{\raisebox{23pt}{ \hspace*{-3pt} $S_z$}}{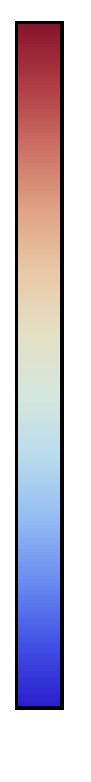}}
\subfigimg[height=4.0cm,angle=0]{\raisebox{-105pt}{ \hspace*{20pt} (d)}}{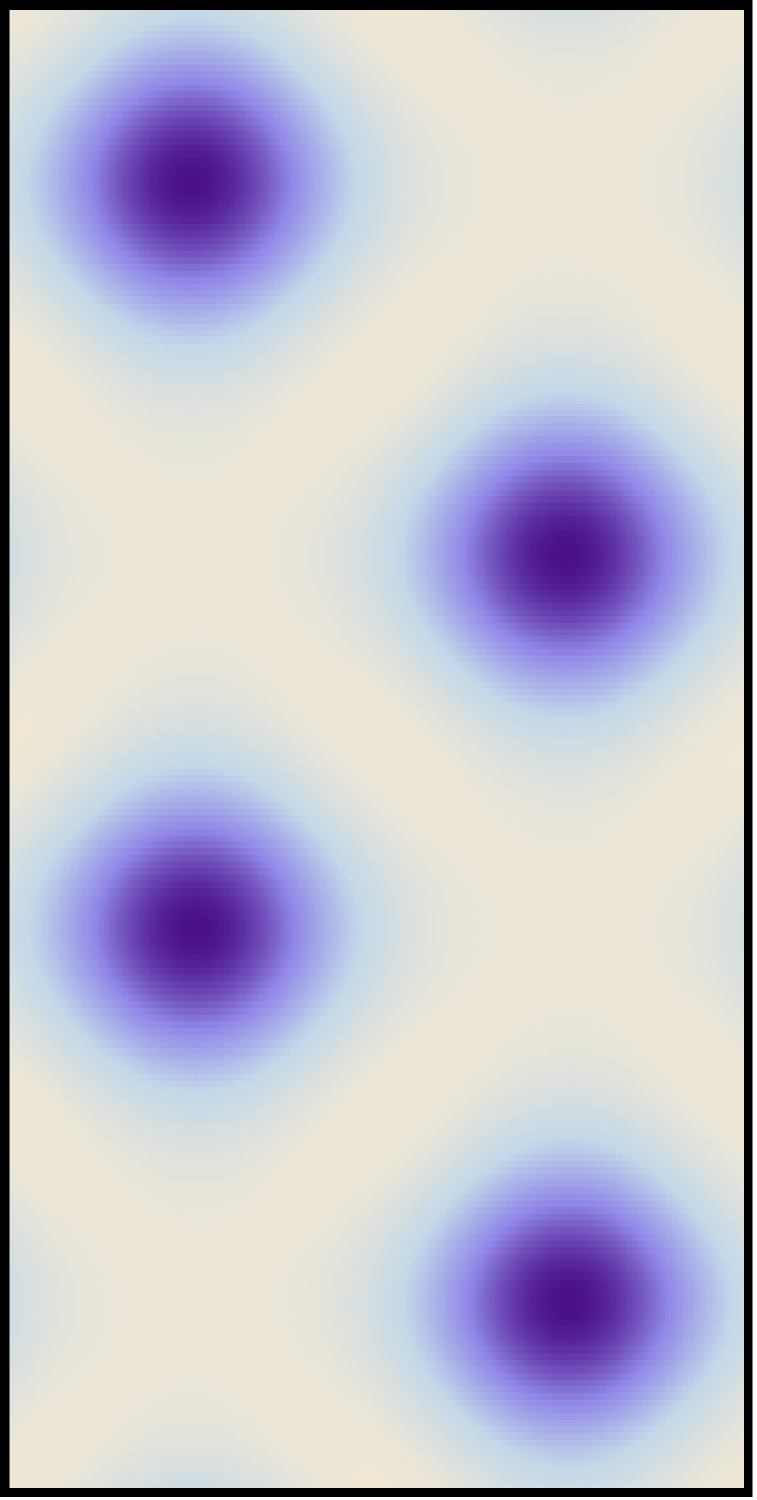}
\subfigimg[height=4.0cm,angle=0]{\raisebox{-105pt}{\hspace*{30pt} (e)}}{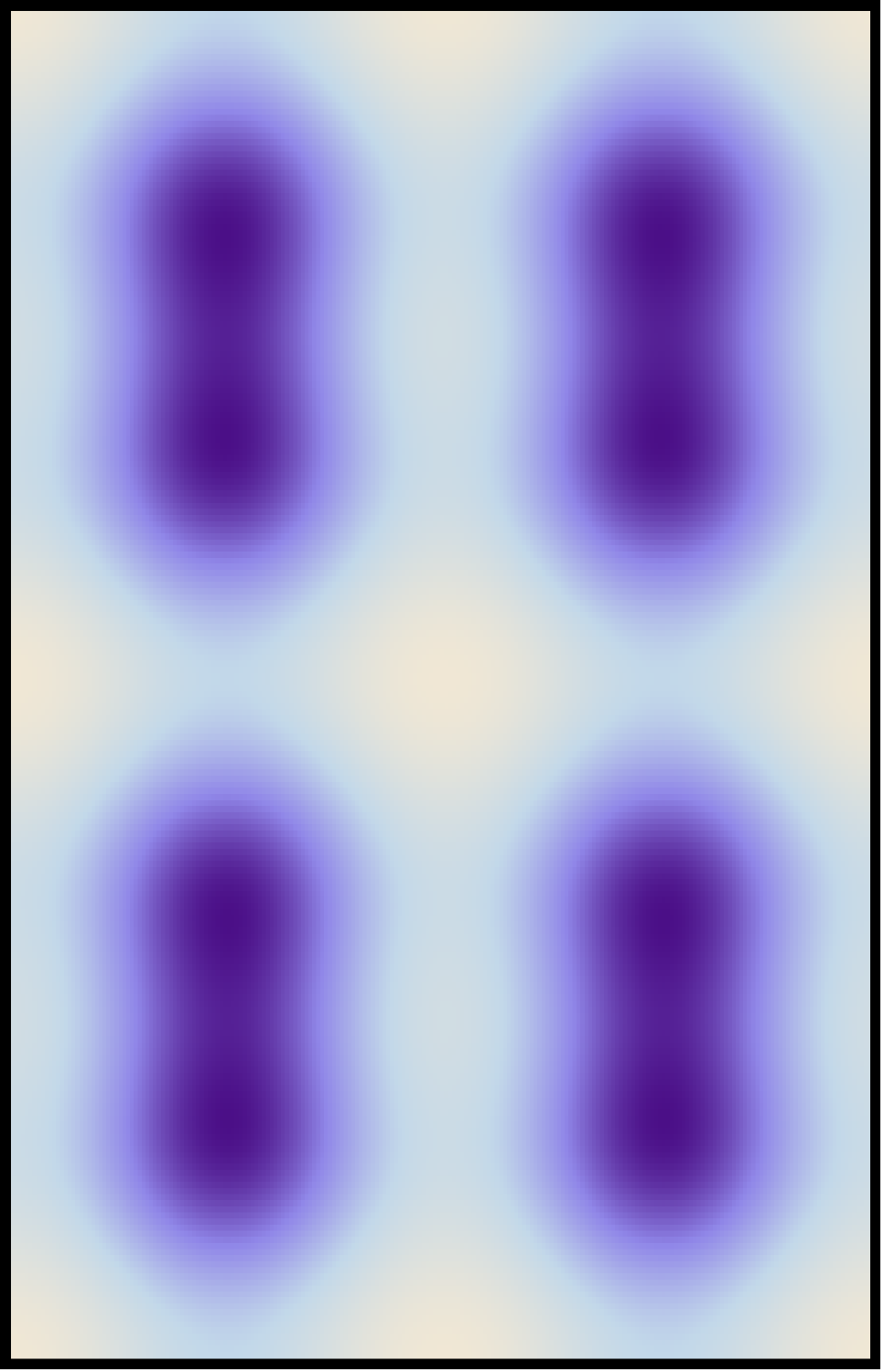}
\subfigimg[height=4.0cm,angle=0]{\raisebox{-105pt}{\hspace*{35pt} (f)}}{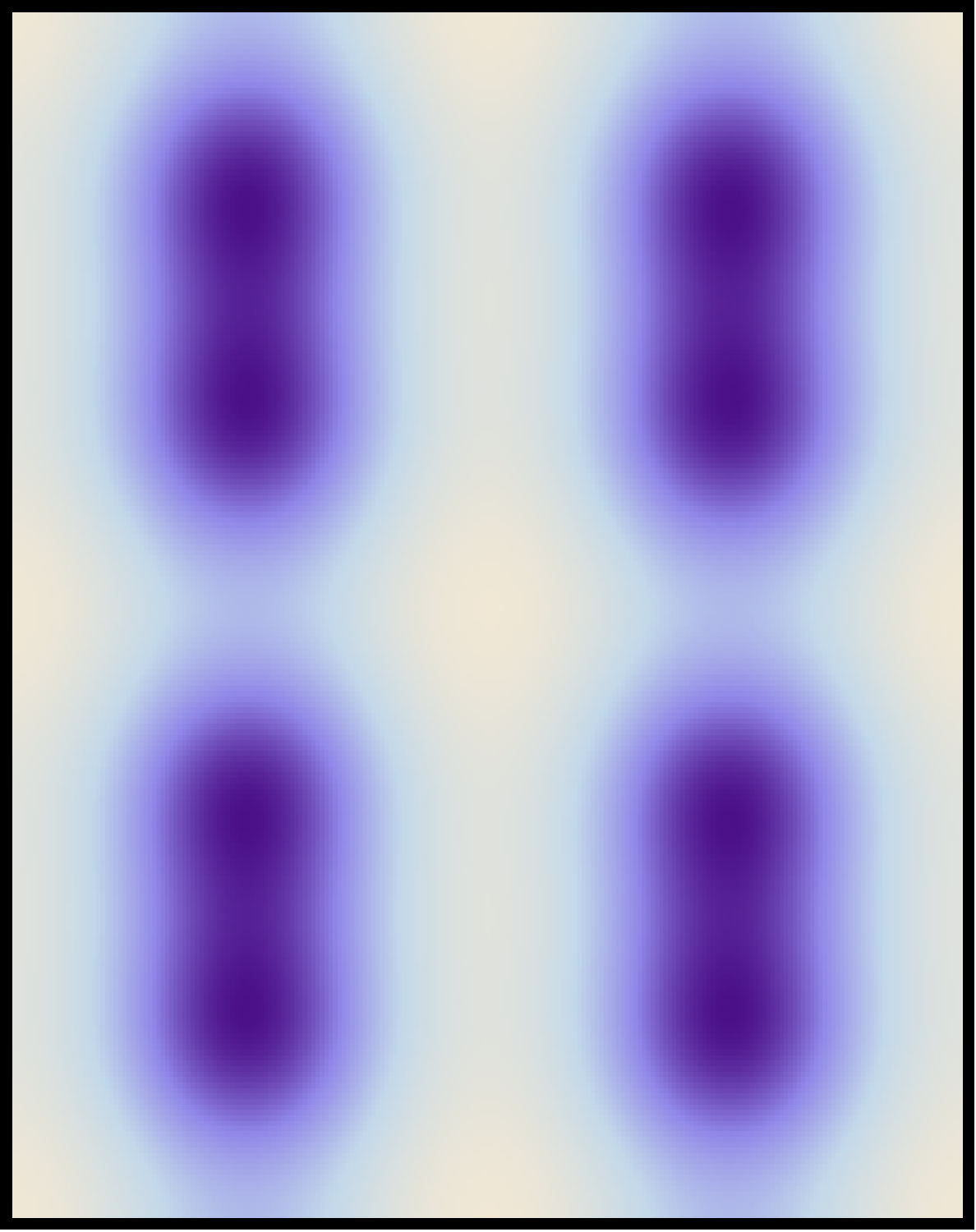}
\hspace*{-2pt}\raisebox{-9.5pt}{\subfigimg[width=0.025\textwidth]{\raisebox{23pt}{ \hspace*{-3pt} $\rho$}}{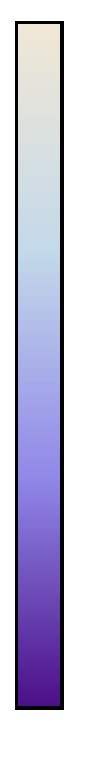}}
\caption{From left to right, we show the total density $\rho$ and $S_z$, for $\Omega_\text{R}/n_v = 0.5,0.25,0.1$ and $\alpha = 1, 0.5, 0.2$, respectively. The aspect ratios are
$\cR = 0.5,0.64,0.79$. For $\alpha = 1$, $S_z$ winds at constant rate around the center of each molecule, whereas for $\alpha <1$
the winding is concentrated around the vortex cores. Consequently, for $\alpha = 1$, the density profile around the molecules is circular (at distances close enough from
the center). On the other hand, for $\alpha <1$ it is elongated.}
\label{fig:5}
\end{figure}

Further, an isolated molecule will have a spin that winds at a constant angular rate around the $x$-axis as we encircle the molecule at fixed radius. It is more natural to regard the molecule as a Skyrmion. Any modulation of the total density is circularly symmetric. For $\alpha\neq 1$ the spin configuration does not wind at a constant rate in the $y-z$ plane, and the total density is anisotropic (see Fig.~\ref{fig:5}).
%

\section{Small Molecule Limit}

The numerical calculations of the previous section show that neighbouring molecules are antialigned. In order to
gain some insight, we develop an effective theory for the system, in the limit of very small molecules.
We assume that the behavior of the lattice of molecules can be explained in terms of the kinetic energy in Eq.~\eqref{E} only.
Within this picture, $g_{ab}$ and $\Omega_\text{R}$ take care of the shape and size of the molecule, but not its orientation.
$\Omega= \pi N_v /A$ ensures the presence of $N_v$ vortices in each component.

\subsection{Kinetic Energy of a Molecule Lattice}

For a system of vortices in 2D with separations $\gg\xi$, the most important contribution to the kinetic energy comes from the regions
away from the vortex cores where $\nabla \sqrt{\rho_{a,b}}=0$ and $\rho_a = \rho_b = n \; (S_z = 0)$. Using the parametrization $\Psi = \sqrt{\rho} e^{-i \chi /2}
(\cos \frac{\theta}{2} e^{-i \varphi /2},\sin \frac{\theta}{2} e^{i \varphi /2})^T$,
\begin{equation}
E_{\text{kin}} = \frac{n}{4} \int d\mathbf{r} \  \left [  (\nabla \chi(\br))^2 + (\nabla \varphi(\br))^2  \right ],
\end{equation}
where $\varphi = \theta_b - \theta_a$ and $\chi = -( \theta_a + \theta_b)$. Note that due to the Rabi field Eq.~\eqref{E_Rabi},
$\varphi =0$ away from the molecules and therefore the only contribution to the kinetic energy away from the
molecule cores comes from the overall phase
\begin{equation}
E_{\text{kin}} = \frac{n}{4} \int d\mathbf{r} \ (\nabla \chi(\br))^2.
\end{equation}
Far from a molecule, $\chi$ obeys Laplaces's equation $\nabla^2\chi=0$, and winds by $4\pi$ as we encircle the molecule, which contains two vortices. Around an isolated molecule with no angular modulation of the density (as occurs at $\alpha=1$, see Fig.\ref{fig:5}) $\chi=2\vartheta$, where $\vartheta$ is the angular coordinate centered on the molecule. In this case, $E_{\text{kin}}$ describes a set of point charges interacting via a 2D Coulomb interactions (see e.g. Ref.~\cite{Kardar}). In the small molecule limit, when $\Omega_\text{R}/g_{ab}n$ becomes large, the kinetic energy dominates the intermolecular energy. If we consider only this contribution, then:

\begin{enumerate}
  \item The molecules form a triangular lattice \cite{Tkachenko:1966}, as confirmed by our numerical calculations.
  \item The orientation of \emph{each} molecule is \emph{separately undetermined}. This corresponds to a lattice of small Skyrmions, each of which may be arbitrarily rotated about the $x$-axis. The freedom will obviously be removed by neglected terms.
\end{enumerate}

For $\alpha<1$ the density modulation around an isolated molecule is angle dependent. Then the angular field $\chi$ of an isolated molecule will have corrections to the point charge configuration that may be described by a multipole expansion, beginning with a quadrupolar field.

The interaction of quadrupoles on a triangular lattice gives rise to a long-ranged aligning interaction between molecules that is absent for $\alpha=1$. Recall that the size of the molecule is determined by $g_{ab}$ and $\Omega_\text{R}$, and is taken
as an input to this picture.

We further assume that in the ground state the net quadrupole moment of the infinite lattice is zero $\sum_i \Delta_i^2=0$. As shown in Appendix~\ref{quadrupole energy}, with this restriction the energy per unit cell
of a periodic lattice with $N_{\text{mol}}$
molecules in the unit cell is given by:
\begin{equation}
\label{E_dens}
V_{\text{inf}}/N_{\text{uc}}= \frac{1}{8} \text{Re} \left [ \sum_{i<j} \Delta_i^2 \Delta_j^2 \mathcal{Q}(z_{ij};\omega_1,\omega_2)   \right ],
\end{equation}
where
\begin{align}
\mathcal{Q}(z_{ij};\omega_1,\omega_2) = &3  \sum_{n,m} \frac{ 1}{(z_{ij}+n \omega_1 + m \omega_2)^4}   \nonumber \\
&-4  \sum_{\substack{n,m\\
                  n^2 + m^2 >0}}  \frac{1}{(n \omega_1 + m \omega_2)^4}    \nonumber \\
                  & -\sum_{k(\neq a)} \sum_{n,m} \frac{ 1}{(z_{ak}+n \omega_1 + m \omega_2)^4}.
\end{align}
The index $a$ corresponds to any molecule in the unit cell. $\mathcal{Q}(z_{ij};\omega_1,\omega_2)$ is
an elliptic function with periods $\omega_1$ and $\omega_2$. We can numerically calculate it truncating the sums.
Here $\omega_1$ and $\omega_2$ shown in Fig.~\ref{fig:6}, are the vectors connecting adjacent unit cells, in complex notation \cite{milne}.

\subsection{Numerical Calculations}

Now that we have an expression for the energy density of the infinite lattice Eq.~\eqref{E_dens}, we can find
the lowest energy configuration in a triangular lattice with the constraint of having zero net quadrupole moment.
We perform a constrained minimization using the SLSQP method implemented in SciPy. We fix the molecules $\Delta_i$ to be
of unit length, so the variables are the orientations of the molecules.

Again, we assume that the infinite system is periodic and therefore we want to find the unit cell. The procedure in this case is
\begin{enumerate}
  \item Minimize the energy Eq.~\eqref{E_dens} with $N_{\text{mol}}$ unit length molecules in the unit cell.
  \item Repeat the minimization, doubling the number of molecules.
  \item If the energy has doubled, we can infer that the unit cell contains $N_{\text{mol}}$ molecules.

\end{enumerate}
Again, this protocol cannot guarantee that there does not exist a larger unit cell with a lower energy density.

\begin{figure}[H]
\centering
\includegraphics[width=5cm]{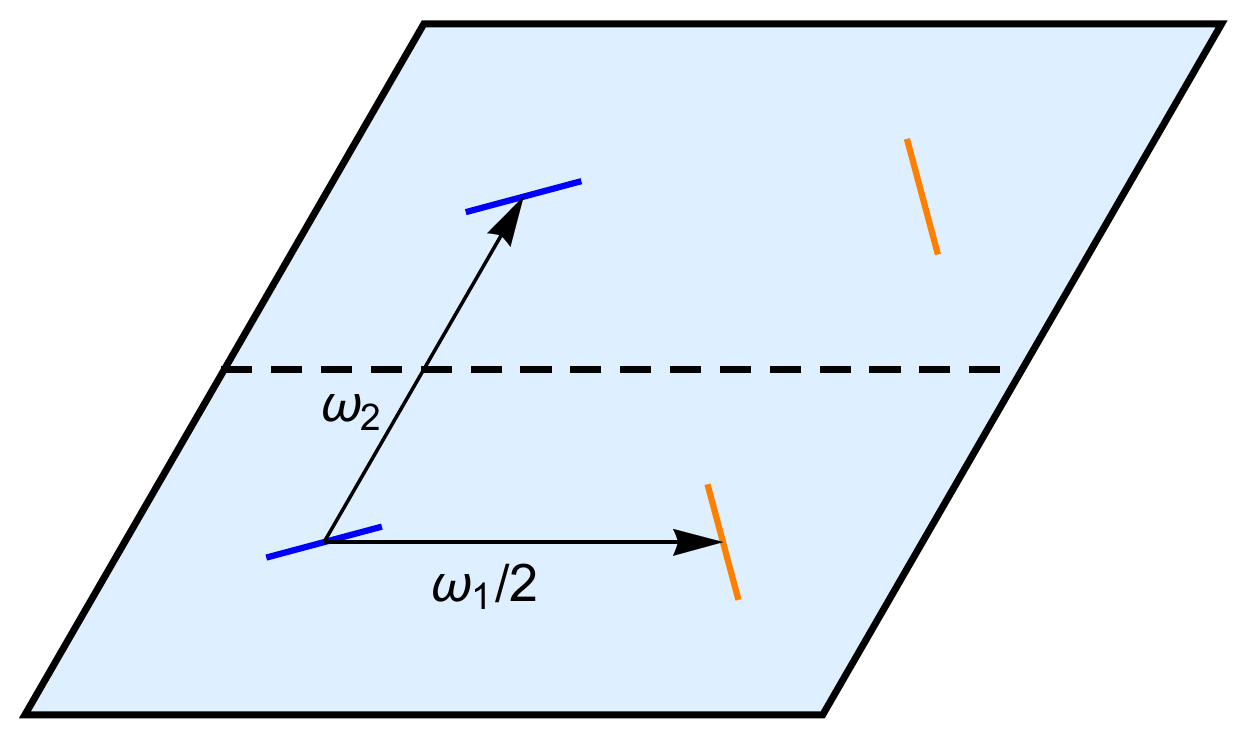}
\includegraphics[width=5cm]{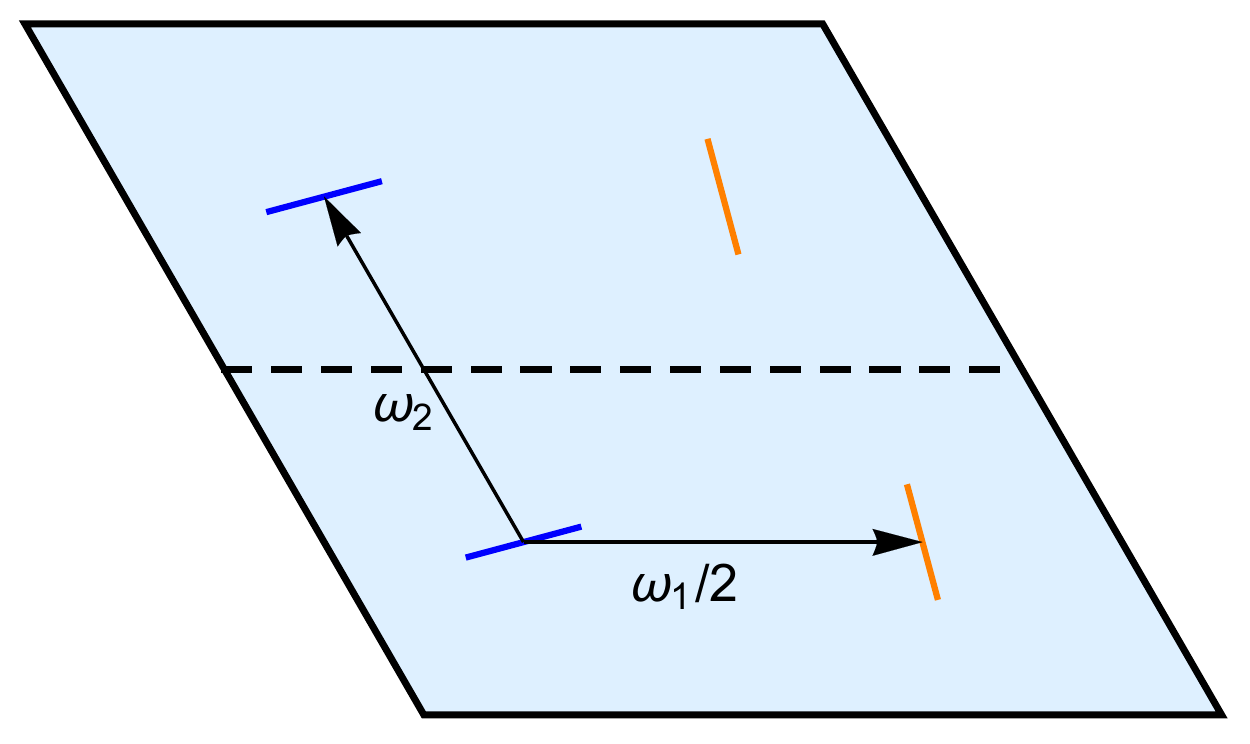}
\caption{We show the two inequivalent configurations with the ground state unit cell we find.}
\label{fig:6}
\end{figure}

We find that the unit cell contains two molecules. Choosing these two molecules to sit on the $x$ axis and $\omega_2 = |\omega_2| e^{i \pi /3}$,
$\mathcal{Q}(z_{12})= |\mathcal{Q}(z_{12})|e^{-i \pi /3}$. Then $V_{\text{inf}}/N_{\text{uc}} = \frac{|\mathcal{Q}(z_{12})|}{8}
\cos (2 \theta_1 + 2 \theta_2 - \pi/3)$. Requiring $e^{i 2 \theta_1}+e^{i 2 \theta_2}=0$, the ground state is
$(\theta_1, \theta_2)= (\frac{\pi}{2}n+\frac{\pi}{12})\pm(0,\frac{\pi}{2})$. There are two inequivalent infinite systems
whose unit cell this is (Fig.~\ref{fig:6}).

\section{Conclusions}

We have calculated the ground state configuration of the infinite system of vortex molecules and found qualitative agreement with previous results in
the bulk of harmonic traps. More interestingly, we have come up with an effective theory in terms of point charges and
derive an expression for the
interaction energy density in terms of an elliptic function $\mathcal{Q}(z_{ij};w_1, w_2)$, in the limit of point vortices and very small molecules when
the net quadrupole moment is zero. We used this expression to find the ground state configuration of an infinite system of
molecules.

\subsection{Acknowledgements}

We are grateful to Luca Mingarelli for very useful discussions. We are grateful to Artane J\'er\'emie Siad for interesting discussions on elliptic functions. BMU and AL would  like  to  acknowledge  EPSRC under  Grants EP/M506485/1 and EP/P034616/1 respectively.

\appendix

\section{Derivation of boundary conditions}
\label{boundary conditions}

Let us parametrize the wave functions as $\psi = \sqrt{\rho}e^{i \theta}$. We define the lattice vectors $\bR_x = L_x \hat{x}$ and $\bR_y = L_y \hat{y}$. $L_x$ and $L_y$ are the dimensions
of the computational unit cell.

We first require the density $\rho_{\sigma}$ in each component to be periodic. This gives us a condition on the amplitudes of the wave-functions:

\begin{align}
 \sqrt{\rho_{\sigma}(\br + \bR_i)} = \sqrt{\rho_{\sigma}(\br)}, \; \sigma = a,b, \; i=x,y.
\end{align}
We next require the superfluid velocity $\bv_{\sigma} = \nabla \theta_{\sigma} - \bA$ on each of the components to be periodic. Note that
$\bv_{\sigma}$ is the $\sigma$ component superfluid velocity only when $\Omega_{\text{R}}=0$. On the other hand, it is still a gauge invariant quantity when $\Omega_{\text{R}} \neq 0$ and hence its periodicity should be required:

\begin{align}
\bv_{\sigma}(\br + \bR_i) = \bv_{\sigma}(\br), \; \sigma = a,b, \; i=x,y.
\end{align}
This does still not give an explicit condition on the phase (it is a set of equations involving gradients of phases). In order to find it, we need to integrate the above equations. By doing so, we arrive to four equations:

\begin{align}
 &\theta_{a}(\br + \bR_x) = \theta_{a}(\br)+ \Omega L_x y + \alpha_x, \\
 &\theta_{a}(\br + \bR_y) = \theta_{a}(\br)- \Omega L_y x + \alpha_y, \\
 &\theta_{b}(\br + \bR_x) = \theta_{b}(\br)+ \Omega L_x y + \beta_x, \\
 &\theta_{b}(\br + \bR_y) = \theta_{b}(\br)- \Omega L_y x + \beta_y.
\end{align}
Hence we are left with four integration constants.

Finally, we require periodicity of spin. Let's define the pseudo-spin density $\bS = \Psi^\dagger \boldsymbol{\sigma}\Psi / 2$, where
$\bm{\sigma}=({\sigma_x, \sigma_y, \sigma_z})$ are the Pauli spin matrices. We first require periodicity of $S_x(\br)$:

\begin{equation}
S_x (\br + \bR_i) = S_x (\br), \; i=x,y.
\end{equation}
These yields the conditions

\begin{align}
 &\beta_x = \alpha_x + 2 \pi n, \\
 &\beta_y = \alpha_y + 2 \pi n, \; \text{$n$ integer.}
\end{align}
This condition also ensures the periodicity of $S_y(\br)$ as well. The periodicity of $S_z(\br)$ is ensured by the periodicity of the densities.
Thus we are left with two integration constants $\alpha_x, \alpha_y$. The only effect of varying these two constants is to shift the wave-functions in the unit cell. These degrees of freedom are expected, since one
reproduces the same infinite lattice upon copying unit cells, no matter where  vortices sit in the computational unit cell.
We choose $\alpha_x = \alpha_y =0$. Note that we have still not said anything about the periodicity of true spin. The key point is that, because $\{ \sigma_0, \sigma_x, \sigma_y, \sigma_z \}$ form a basis in the space of $2 \times 2$ matrices, periodicity
of $\mathbf{S}(\br)$ automatically ensures periodicity of spin. Hence, in order to fully fix the boundary condition for the wave-functions, it is necessary and sufficient to require
periodicity of $\rho_{\sigma}, \bv_{\sigma}$ and $\bS$. The boundary condition is:

\begin{align}
\label{BC}
 &\psi_{\sigma}(\br + \bR_x) = e^{i \Omega L_x y} \psi_{\sigma}(\br) \nonumber \\
  &\psi_{\sigma}(\br + \bR_y) = e^{-i \Omega L_y x} \psi_{\sigma}(\br) , \; \sigma = a,b.
\end{align}

\section{$\Omega$: allowed values and relation to the number of vortices}
\label{allowed values and relation to the number of vortices}

Let's prove that in order to avoid having a contradicting theory, the angular velocity can only take a discrete set of values. Using Eq.~\eqref{BC}:

\begin{align}
 \psi(x+L_x , y+L_y) &= e^{-i \Omega L_y (x+L_x)} \psi(x+L_x,y)  \nonumber \\
 &= e^{-i \Omega L_y (x+L_x)} e^{i \Omega L_x y} \psi(x,y) \nonumber \\
 \psi(x+L_x , y+L_y) &=e^{-i \Omega L_x (y+L_y)} \psi(x,y+L_y) \nonumber \\
 &= e^{i \Omega L_x (y+L_y)} e^{-i \Omega L_y x} \psi(x,y)
\end{align}

\begin{equation}
\label{omega}
 \psi(x+L_x , y+L_y) = \psi(x+L_x , y+L_y) \Leftrightarrow \Omega = \frac{\pi n}{A},
\end{equation}
where $n$ is an integer and $A=L_x L_y$.

In his seminal work, Feynman explains that the lowest energy state for an irrotational fluid with a given angular momentum is a vortex lattice, with a $2 \pi$ winding of the phase around each vortex \cite{Feynman:1955aa}. Because the superfluid velocity
is $\bv = \nabla \theta$, the superfluid cannot rotate as a rigid body. On the other hand, the vortex lattice (the set of vortex cores) can only
rotate as a rigid body in equilibrium \cite{Tkachenko:1966}. Hence, on average the region of the superfluid that is packed with vortices, rotates as a rigid body. This allows to estimate a relation between the angular velocity of the trap $\Omega$ and the number
of vortices $N_v$, in the ground state. Let $D$ be a region of area $A$ packed with $N_v$ vortices and $\bv = \Omega r \hat{\varphi}$ (rigid solid rotation). Let $\partial D$ be its boundary. If we calculate the
circulation of the velocity:

\newenvironment{rcases}
  {\left.\begin{aligned}}
  {\end{aligned}\right\rbrace}

\begin{equation*}
\begin{rcases}
  \Gamma_{\partial D} = \oint_{\partial D} \bv \cdot d \bl = 2 \Omega A \\
  \Gamma_{\partial D} = 2 \pi N_v
\end{rcases}
 \Rightarrow \Omega = \frac{\pi N_v}{A}.
\end{equation*}
From this result, we infer the meaning of $n$ in Eq.~\eqref{omega}: $n=N_v$. Even though this estimate is based on heuristic arguments, it is verified in the numerics, i.e. the number of vortices found in the ground state is $n$.

\section{Numerical integration}
\label{Numerical integration}

We have an integral in two dimensions over the area $A$

\begin{equation}
 E = \int_{A} d \mathbf{r} \mathcal{E}(\mathbf{r}).
\end{equation}
To calculate the integral numerically, we use the $Riemann \; sum$ method. We have a 2D grid
of points $(x_n , y_m)$, where $x_n = n a_x$, $n,m \in [0, N -1]$, $y_n = n a_y$. The discretization
lattice constant is $a_x = L_x /N$ and $a_y = L_y /N$. Using this method, we approximate the integral
as the sum of the  volumes of the parallelepipeds:

\begin{equation}
 V_{nm} = a_x a_y \mathcal{E}(x_n,y_m),
\end{equation}
and

\begin{equation}
  E(\text{estim.}) = a_x a_y \sum_{n,m} \mathcal{E}(x_n,y_m).
\end{equation}
In an analogous way to what is done in \cite{riley}, we find that to leading order, the discretization error is

\begin{equation}
\label{error}
 \Delta E = E(\text{estim.}) - E = - \frac{A}{2}(a_x \left \langle \partial_x \mathcal{E} \right \rangle + a_y \left \langle \partial_y \mathcal{E} \right \rangle),
\end{equation}
where $ \left \langle \partial_x \mathcal{E} \right \rangle = \frac{1}{N^2} \sum_{nm} \partial_x \mathcal{E}(x_n + a_x / 2, y_m + a_y /2)$.

Let's now write the discrete version of Eq.~\eqref{E}, with $\omega_{\text{eff}}=0$. It is all trivial to write except the kinetic energy term, so let's focus on that first. We introduce the vector potential by
making the Peierls substitution:

\begin{align}
&\psi^{*}(x)(p_x-A_x)^2 \psi(x) = \nonumber \\
&\frac{1}{a^2} \big ( 2|\psi(x)|^2-e^{iA_x a}\psi^{*}(x+a)\psi(x)-e^{-iA_x a}\psi^{*}(x)\psi(x+a) \big )\nonumber \\
&+ \cO (a^3).
\end{align}
Now, by defining the discrete wave function $\varphi(n,m)=\psi(x,y)/\sqrt{a_x a_y}$:

\begin{align}
\label{E_d}
 &E_d(\cR,\{\varphi_{\sigma}(n,m) \}) = \nonumber \\
& \sum_{\sigma}\sum_{n,m}  \Big [  \frac{1}{2a_{x}(\cR)^{2}}\left|\varphi_{\sigma}(n+1,m)e^{-iA_x a_x(\cR)}
-\varphi_{\sigma}(n,m)\right|^2  \nonumber \\
&  +\frac{1}{2a_{y}(\cR)^{2}}\left|\varphi_{\sigma}(n,m+1)e^{-iA_y a_y(\cR)}-\varphi_{\sigma}(n,m)\right|^2  \nonumber \\
&   - \mu_{\sigma} |\varphi_{\sigma}(n,m)|^2  \Big ] \nonumber \\
 &+ \sum_{\sigma_1 , \sigma_2} \frac{g_{\sigma_1 \sigma_2}}{2} \sum_{n,m} \frac{1}{a_x(\cR) a_y(\cR)}  |\varphi_{\sigma_1}(n,m)|^2 |\varphi_{\sigma_2}(n,m)|^2 \nonumber \\
 & -\Omega_\text{R}\sum_{n,m} \left [\varphi_{a}^{*}(n,m) \varphi_{b}(n,m) +\varphi_{b}^{*}(n,m) \varphi_{a}(n,m) \right ].
\end{align}

Note that we have added two Lagrange multipliers $\{ \mu_{a} , \mu_{b}\}$ to constrain the norm of the wave function (and therefore the number of particles) in the minimization.

Here something important pointed out by Mingarelli $et \; al$ is that in order to allow the vortex lattice configuration access any lattice geometry, one needs to minimize with respect to the aspect ratio
of the unit cell $\cR = a_x/a_y$. It is important to parametrize the discretization lattice constants in such a way that the area of the computational unit cell does not depend on it. Such a parametrization is:

\begin{equation}
\label{a(R)}
 a_x(\cR) = \sqrt{\frac{A}{N^2} \mathcal{R}} \;\; \text{and} \;\; a_y(\cR) = \sqrt{\frac{A}{N^2} \frac{1}{\mathcal{R}}}.
\end{equation}

\section{Energy of the infinite system of point charge molecules}
\label{quadrupole energy}

If $V_{ij}$ is the interacting energy of a pair of molecules, the energy of the infinite lattice is:

\begin{align}
 V_{\text{inf}} &= \frac{1}{2} \sum_{i \neq j} V_{ij} = \frac{1}{2} \sum_{i=1}^{N_{\text{mol}}} \sum_{j(\neq i)} V_{ij} = \nonumber \\
 &\frac{1}{2} \sum_{i=1}^{N_{\text{uc}}} \sum_{i'=1}^{N_{\text{mol/uc}}} \sum_{j(\neq k)} V_{kj},
\end{align}
where $k = (i-1)N_{\text{uc}}+i'$, $N_{\text{mol}}=\infty$, $N_{\text{mol/uc}}$ is the number of molecules in the unit cell and $N_{\text{uc}}$ is the
number of unit cells, assuming there is some periodicity in the infinite system. Now, assuming so, i.e. assuming $V_{ij}$ depends only on
the vector $\br_{ij}$ and not $\br_i$ and $\br_j$ separately and assuming the unit cell contains $N_{\text{mol}}$ molecules, we can write the energy per unit cell:

\begin{equation}
\label{V/N}
 V_{\text{inf}}/N_{\text{uc}}= \frac{1}{2} \sum_{i=1}^{N_{\text{mol/uc}}} \sum_{j(\neq i)}V_{ij}.
\end{equation}
Here $j$ runs over the infinite set of molecules.

Let's now write the specific expression for the energy. Because it is a 2D problem, it is convenient to use complex numbers instead of
vectors \cite{milne}. The 2D Coulomb potential energy of two charges sitting at $z_1$ and $z_2$ is $- \text{Re} \log (z_1 - z_2)$ \cite{Kardar}.

Let's now consider a pair of molecules, with repelling charges at
\begin{align}
	&z_{1,2}^a = \zeta_{1,2}+ \frac{\Delta_{1,2}}{2}, \\
	&z_{1,2}^b = \zeta_{1,2}- \frac{\Delta_{1,2}}{2}.
\end{align}
With the molecule lengths $|\Delta_1|=|\Delta_2|<< |\zeta_{12}|$, where $\zeta_{12}=\zeta_1-\zeta_2$. Ignoring the interaction
within a molecule, the interaction energy is proportional to the real part of
\begin{align}
\label{multipole}
	&-\log\left[(z_1^a-z_2^b)(z_1^b-z_2^a)(z_1^a-z_2^a)(z_1^b-z_2^b)\right] = \nonumber \\
	&-4 \log \zeta_{12}- \log\left[ 1- \left (  \frac{\Delta_1+\Delta_2}{2\zeta_{12}} \right )^2 \right ]- \nonumber \\
	&\log\left[ 1- \left (  \frac{\Delta_1-\Delta_2}{2\zeta_{12}} \right )^2 \right ] \nonumber \\
	&= -4 \log \zeta_{12} + \frac{\Delta_1^2 + \Delta_2^2}{2 \zeta_{12}^2} +\frac{\Delta_1^4 + \Delta_2^4 + 6\Delta_1^2 \Delta_2^2 }{(2 \zeta_{12})^4} \nonumber \\
	 &+ \mathcal{O}((\Delta_1 / \zeta_{12})^6).
\end{align}
Let's now calculate the energy per unit cell. The first term in the expansion is addressed by putting the molecules in a triangular lattice. Let's start by calculating the next to leading order contribution.  We will derive an expression for the case of arbitrary number of molecules in the unit cell and for an arbitrary periodic lattice geometry. In order to follow the calculations let's illustrate one concrete case:

\begin{figure}[H]
\centering
\includegraphics[width=7cm]{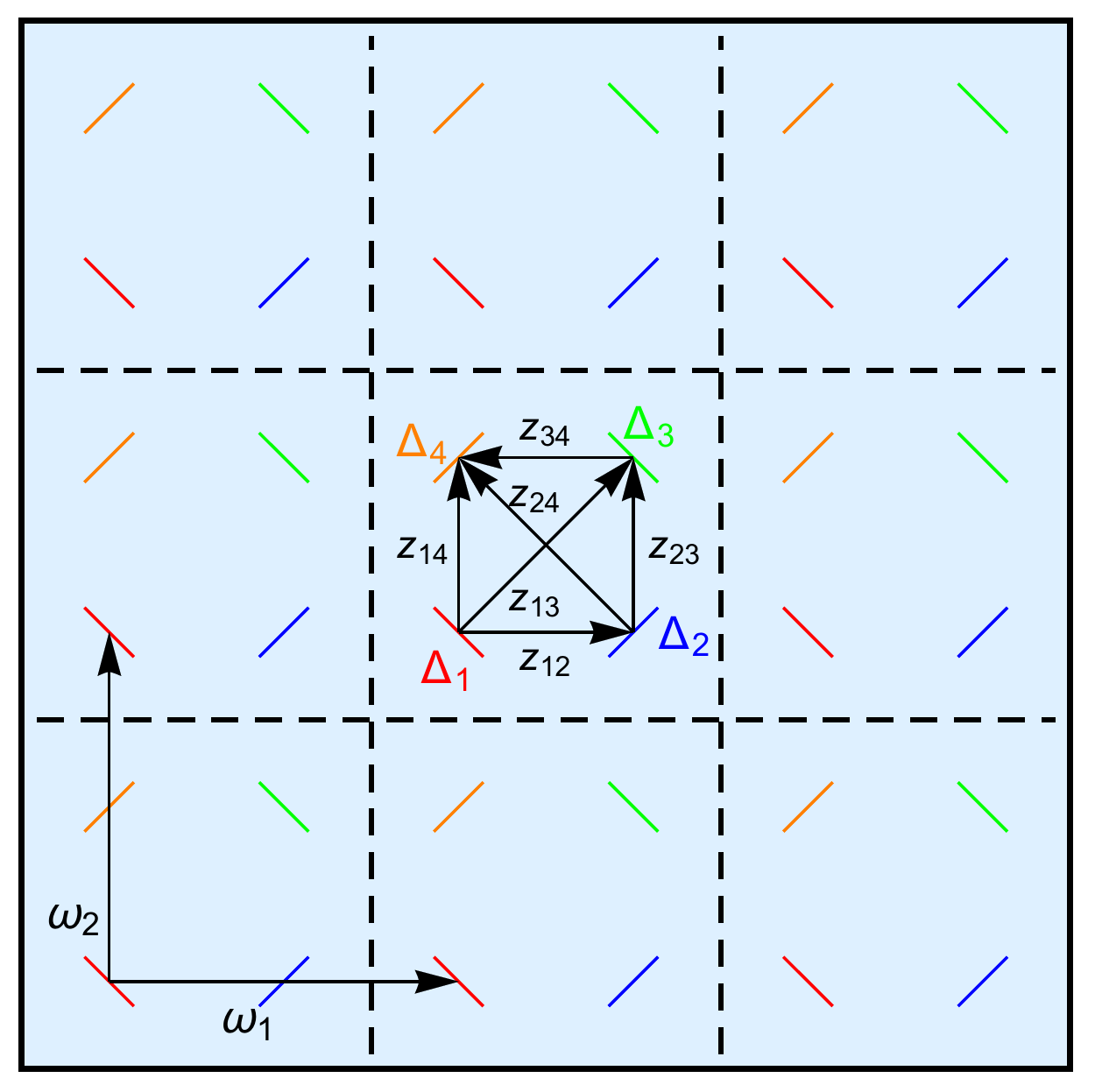}
\caption{A schematic square lattice of molecules with $N_{\text{mol}}=4$. Even though our numerical calculations are done for the triangular lattice, we show a square lattice
to emphasise that our result of the general expression of $V_{\text{inf}}/N_{\text{uc}}$ is valid for any lattice geometry.}
\label{dip_lattice}
\end{figure}
We need to sum the interaction energy of each of the molecules in only one unit cell, with all the other molecules in the infinite lattice. Let's start
with the contribution coming from pairs, where one of the molecules is outside of the unit cell. Let $\omega_1$ and $\omega_2$ be the vectors connecting adjacent unit cells and $\{ z_{ij}\}$ the vectors connecting molecules within a unit cell,
in complex notation \cite{milne}. We first do equal color (i.e. equal position in the unit cell) pairs, each of them contributes:

\begin{equation}
 \sum_{\substack{n,m\\
                  n^2 + m^2 >0}} \frac{(\Delta_i^2 + \Delta_i^2 )}{2(n \omega_1 + m \omega_2)^2}.
\end{equation}
Here and from now on, $n,m$ run over all the integers that fulfill the condition $n^2 + m^2 > 0$ (if specified). The indices
$i,j$ and $k$ run over molecules inside the unit cell. Therefore, the total contribution is:

\begin{equation}
\label{eq}
 \sum_{i=1}^{N_{\text{mol}}}   \sum_{\substack{n,m\\
                  n^2 + m^2 >0}} \frac{(\Delta_i^2 + \Delta_i^2 )}{2(n \omega_1 + m \omega_2)^2}.
\end{equation}
The contribution from different color pairs is only slightly trickier, $-z_{ij}$ appears for some of the terms in the denominator, but because of the
square, the signs disappear. The contribution of each of the members of the unit cell is:

\begin{equation}
\sum_{j=1}^{N_{\text{mol}}} \sum_{\substack{n,m\\
                  n^2 + m^2 >0}} \frac{ (\Delta_i^2 + \Delta_j^2 ) }{2(z_{ij}+n \omega_1 + m \omega_2)^2}.
\end{equation}
Therefore the total contribution is:

\begin{equation}
\label{neq}
\sum_{i \neq j}^{N_{\text{mol}}} \sum_{\substack{n,m\\
                  n^2 + m^2 >0}} \frac{ (\Delta_i^2 + \Delta_j^2 )}{2(z_{ij}+n \omega_1 + m \omega_2)^2}.
\end{equation}
The contribution from pairs inside the unit cell is just:

\begin{equation}
\label{intra}
\sum_{i \neq j}^{N_{\text{mol}}} \frac{(\Delta_i^2 + \Delta_j^2 )}{2z_{ij}^{2}}.
\end{equation}

Summing all contributions Eq.~\eqref{eq}, Eq.~\eqref{neq} and Eq.~\eqref{intra} and including the factor $1/2$ from Eq.~\eqref{V/N}, we get the final expression:

\begin{align}
 &V_{\text{inf}}/N_{\text{uc}}= \frac{1}{2} \text{Re} \Bigg [ \sum_{i }^{N_{\text{mol}}} \Delta_i^2  \sum_{\substack{n,m\\
                  n^2 + m^2 >0}}  \frac{1}{(n \omega_1 + m \omega_2)^2} + \nonumber \\
  & \sum_{i<j} (\Delta_i^2 + \Delta_j^2)  \sum_{n,m}
                    \frac{ 1}{(z_{ij}+n \omega_1 + m \omega_2)^2} \Bigg ].
\end{align}
The first term is divergent unless we require the net quadrupole moment of the unit cell to vanish $\sum_i \Delta_i^2 =0$. Using this condition and due to the double periodicity in $z_{ij}$ of the infinite sum in the second term, $V_{\text{inf}}/N_{\text{uc}}=0$.

We next calculate the contribution of the third term in the multipole expansion Eq.~\eqref{multipole}. Following the same steps as above we
arrive to
\begin{align}
&V_{\text{inf}}/N_{\text{uc}}= \frac{1}{2^4} \text{Re} \Bigg [ 4 \sum_i \Delta_i^4   \sum_{\substack{n,m\\
                  n^2 + m^2 >0}}  \frac{1}{(n \omega_1 + m \omega_2)^4} + \nonumber \\
                  & \sum_{i<j} (\Delta_i^4 + \Delta_j^4 + 6 \Delta_i^2 \Delta_j^2) \sum_{n,m}
                    \frac{ 1}{(z_{ij}+n \omega_1 + m \omega_2)^4} \Bigg ].
\end{align}
Using $\sum_i \Delta_i^2 =0$ $(\Rightarrow \sum_i \Delta_i^4 = - 2\sum_{i<j} \Delta_i^2 \Delta_j^2)$ and taking advantage
of the double periodicity in $z_{ij}$  of the last infinite sum, we can rewrite the expression as
\begin{equation}
V_{\text{inf}}/N_{\text{uc}}= \frac{1}{8} \text{Re} \left [ \sum_{i<j} \Delta_i^2 \Delta_j^2 \mathcal{Q}(z_{ij};\omega_1,\omega_2)   \right ],
\end{equation}
where
\begin{align}
\mathcal{Q}(z_{ij};\omega_1,\omega_2) = &3  \sum_{n,m} \frac{ 1}{(z_{ij}+n \omega_1 + m \omega_2)^4}   \nonumber \\
&-4  \sum_{\substack{n,m\\
                  n^2 + m^2 >0}}  \frac{1}{(n \omega_1 + m \omega_2)^4}    \nonumber \\
                  & -\sum_{k(\neq a)} \sum_{n,m} \frac{ 1}{(z_{ak}+n \omega_1 + m \omega_2)^4}.
\end{align}
Here $a$ is any molecule in the unit cell. $\mathcal{Q}(z_{ij};\omega_1,\omega_2)$ is an elliptic function with periods $\omega_1$
and $\omega_2$. We can numerically calculate it truncating the sums.

\bibliographystyle{apsrev4-1}
\bibliography{biblio_moleculelattice.bib}

\end{document}